\documentclass[iop]{emulateapj}

\shorttitle{Spatial Nonlocality of the Small-Scale Solar Dynamo}
\shortauthors{Lamb et~al.}

\usepackage{rotfloat}

\begin{document}

\title{Spatial Nonlocality of the Small-Scale Solar Dynamo}

\author{D.~A. Lamb, T.~A. Howard, and C.~E. DeForest}

\affil{Southwest Research Institute,
  1050 Walnut Street, Suite 300, Boulder, CO, 80302, USA}
\email{derek@boulder.swri.edu}

\begin{abstract}
We explore the nature of the small-scale solar dynamo by
tracking magnetic features. We investigate two
previously-explored categories of the small-scale
solar dynamo: shallow and deep. Recent modeling work on the shallow
dynamo has produced a number of scenarios for how a strong network
concentration can influence the formation and polarity of nearby
small-scale magnetic features. These scenarios have measurable
signatures, which we test for here using magnetograms
  from the Narrowband Filter Imager (NFI) on \emph{Hinode}. We find
no statistical tendency for newly-formed magnetic features to cluster
around or away from network concentrations, nor do we find any
statistical relationship between their polarities. We conclude that
there is no shallow or ``surface'' dynamo on the
spatial scales observable by \emph{Hinode}/NFI. In light of these
results, we offer a scenario in which the sub-surface field in a deep
solar dynamo is stretched and distorted via turbulence,
allowing the field to emerge at random locations on
the photosphere.
\end{abstract}


\section{Introduction}
\label{sec:intro}

The Sun is covered in a pattern of ``salt and pepper'' small magnetic
features that dominate the star's photospheric magnetic energy budget
\citep{Babcock1953}.  The generation of large-scale magnetic fields in
the interior and on the surface of the Sun is understood to be a
consequence of rotational motion, large-scale convection, the
transport of magnetic fields to the poles, and the storage of
intensely strong magnetic fields at the base of the convection zone,
though many important questions remain unanswered
\citep{Charbonneau2010}. The origin of magnetic fields on much smaller
spatial scales, scales of order the supergranular ($\sim$15--30 Mm)
size, granular ($\sim$1 Mm) size, or smaller, is not as well
understood.  These small-scale magnetic fields are important for the
overall magnetic flux and energy budgets of the Sun, and are important
in structuring and heating the chromosphere and corona.

On one hand, it is possible that the fields seen on these small scales
are produced as a consequence of the global, large scale (mean field)
dynamo. In this scenario, the convective buffeting of the fields at
the surface and in the interior shreds them to smaller and smaller
scales, down to (and likely further than) the resolution limit of
currently available telescopes.  Evidence of this model is given by
simulations of magnetoconvection that show invariance across a wide
range of scales \citep{SteinNordlund2006} and by observations that the
probability distribution of magnetic flux concentrations shows a
smooth $-1.8$ power-law distribution over nearly 6 orders of magnitude
in magnetic flux \citep{Parnell2009}.

On the other hand, it is possible that a shallow near-surface
``small-scale'' dynamo is present, and that even without the global dynamo,
magnetic fields would continue to be generated and amplified by the
small-scale flows. Recent observational evidence, primarily from
\emph{Hinode}, lends some credence to this scenario. Examples include
the patterns of horizontal magnetic fields in the photosphere
\citep{Lites2008} and the lack of a change in the numbers of weak
magnetic features in the quiet sun when measured as a function of the
solar cycle phase \citep{Buehler2013}.

Several groups have produced impressively realistic-looking
simulations that amplify a small seed field into something that looks
and behaves like the observed magnetic network
\citep[e.g.,][]{Cattaneo1999, VoglerSchussler2007}. However, as
\cite{Stenflo2012} recently pointed out, evidence for
small-scale dynamo activity in these simulations is
not necessarily evidence for small-scale dynamo
activity on the Sun, as the results typically depend strongly on the
initial conditions or the approximations used. Due to unavoidable
computational limitations, current state-of-the-art simulations are
forced to operate in a physical regime in which the Reynolds number
$Re$, magnetic Reynolds number $Re_M$, and magnetic Prandtl number
$Pm$ are (sometimes vastly) dissimilar to the actual properties of the
Sun. Thus while the simulations are useful for understanding the size
scales, expected magnetic phenomenology, and interaction of any
generated fields with larger-scale
fields, progress in understanding the existence of and role played by
a small-scale dynamo is, for now, best made by
observational analysis.

There is a third alternative to the problem of the
  small-scale flux: a small-scale dynamo in which the dual processes
  of stretching of the seed field and the addition of the new field to
  the photosphere are not in close proximity to each other-the new
  field may be observed at an essentially random location with respect
  to the original field. Such a dynamo could be driven by turbulent
  convection throughout the full depth of the convection zone without
  the proximity properties one would expect of a shallow surface
  dynamo. This possibility is supported by simulations that show cool
  downdrafts extending through the turbulence of the outer solar
  layers to the base of the convection zone
  \citep[e.g.,][]{Stein2003}. If this possibility is correct the
  cross-scale equilibrium \citep{Schrijver1997} could drive energy
  flow in either direction: large to small scales, or vice versa
  \citep[e.g.,][]{VoglerSchussler2007}. We refer to this possibility
  as a ``spatially nonlocal small-scale dynamo'', where ``spatially''
  is meant to distinguish (non-)locality in physical space from
  (non-)locality in wavenumber space.  

Dynamos are a mechanism for creating magnetic fields, but in order for
a dynamo to operate indefinitely, there must be a way to stem the
continued growth of the field. Otherwise, the photosphere would
quickly become choked with magnetic fields and convection suppressed, which is obviously not the
case as can be seen in any photospheric line-of-sight magnetogram.
Very generally, this stemming of the growth of the field may take two
forms: in the first, existing magnetic field is removed in
cancellation / flux annihilation events; in the second, the production
of new flux is suppressed.  While the cancellation events have been
studied for decades, the suppression of continued generation of
magnetic flux (apart from sunspots) is less extensively studied.  Two
relatively recent examples include the work of \cite{Morinaga2008}, in
which convection (and thus presumably dynamo action) was reduced in
the presence of small-scale magnetic features such as G-band bright
points, and the work of \cite{Hagenaar2008}, in which the emergence
rate of large-scale ephemeral regions was found to be reduced in areas
of strong unipolar magnetic fields (including but not limited to
coronal holes).

\subsection{Outline}
\label{sec:Outline}

In this paper, we bridge the gap between the works of
\cite{Morinaga2008} and \cite{Hagenaar2008} and investigate the
spatial relationship between existing strong-field regions and the
detection of new magnetic features, which we take as a proxy for
dynamo action, at intermediate spatial scales.  We focus on the areas
around supergranular network flux concentrations, and analyze whether
the rate of feature birth at a range of distances from the network
concentration is significantly larger or smaller than would be
expected from a random distribution of events.

In addition to occupying an interesting intermediate spatial scale,
supergranular network concentrations are an ideal observational
target: their field strength and flux is high enough that they might
reasonably have a positive or negative effect on any
spatially local small-scale dynamo activity, their
occurrence is common enough that several will exist in a
reasonably-sized dataset, and they are sufficiently long-lived such
that the surrounding plasma can be affected by their presence. Our
criteria for identifying network concentrations is given in
Section~\ref{sec:feature-tracking}. Our goal was to identify whether
the rate of detection of new magnetic features has some dependence on
the distance from the network concentration.

In this work we use the number of features as a proxy for the rate of
flux production. This enables us to explore four mechanisms by which
the number of features at short distances from a network concentration
could be altered. In the case of suppression, for example, fewer
features should be found near the borders of the concentration, while
more features per unit area should be found further away.

\begin{figure}
\includegraphics[width=1.0\columnwidth]{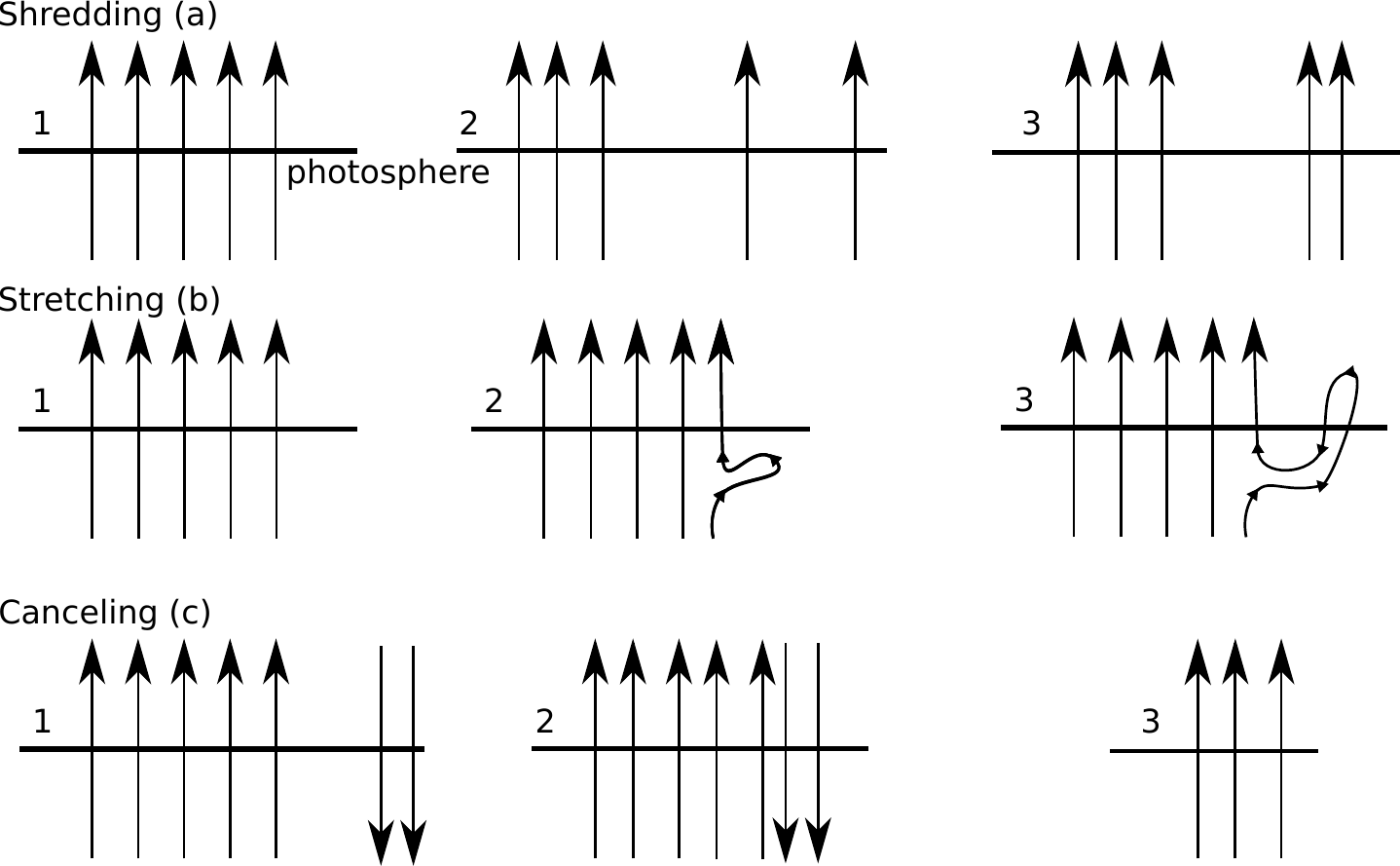}
\caption{Three mechanisms which could result in an increase in detected 
features near a network concentration. a) As the network concentration 
is shredded, flux is removed from the edge of the concentration, possibly 
(but not necessarily) in a way that eludes detection, as shown. b) Network 
concentration field lines below the surface are stretched and brought to the 
surface. c) Opposite polarity flux impinges on the network concentration and 
cancels.}
\label{fig:Shred-&-stretch}
\end{figure}

Three forms of feature evolution are illustrated in
Figure~\ref{fig:Shred-&-stretch}.  Shredding
(Figure~\ref{fig:Shred-&-stretch}a) involves field lines that are
bodily moved away from the network concentration. This results in a
decrease of the concentration's flux as the field lines move away, and
the features have the same polarity as the network
concentration. Stretching (Figure~\ref{fig:Shred-&-stretch}b) involves
field lines from below the surface that are stretched and brought to
the surface.  Such distorted field components would then appear on the
Sun as newly-formed small-scale features.  Here the flux of the
concentration remains the same, the total unsigned flux in the region
increases, and the new, nearby features have a mixed polarity. This is
the chief mechanism we search for as evidence of spatially local small-scale dynamo
action. Canceling (Figure~\ref{fig:Shred-&-stretch}c) involves field
lines of opposite polarity to the network concentration that are
brought towards it and cancel with it. In this case the flux of the
network concentration decreases and the unsigned flux of the region
also decreases.

Section~\ref{sec:Methodology} describes the dataset and feature
tracking method, the means for identifying network concentrations, and
the method used to measure suppressions and enhancements of features
around the network concentrations. Section~\ref{sec:results} presents
our results. We find no signature of systematic enhancement or
suppression of magnetic feature production in the vicinity of the
network concentrations. Instead, we find that most measured
suppressions and enhancements can be attributed to network
concentration evolution, as shown in
Section~\ref{sub:NetworkConcEvolution}. Further, we find that there is
no statistical tendency for these magnetic features to cluster around
network concentrations. Section~\ref{sec:Clustering-Discussion}
concludes with a discussion on the implications of these null
measurements and the importance of small-scale fields to network
concentration evolution.

\section{Methodology}
\label{sec:Methodology}

\subsection{Data Processing }
\label{sec:Observations}

We use line-of-sight magnetograms from the \emph{Hinode}/NFI
instrument \citep{Kosugi2007,Tsuneta2008}.  The dataset and its
preparation was described in detail by \citet{Lamb2010}. Briefly, the
data consist of a 5.25~h long sequence of 420 $277"\times 96"$ (0.16''
pixels) Na D 5896~\AA\ magnetograms with a cadence of 45~s, beginning
at 2007-09-19 12:44:44~UT and ending at 17:58:59~UT.
The observed region was just south of disk center, in
  a region of quiet sun, and Hinode tracked the region as it rotated
  across the disk.  The magnetograms were converted from raw Stokes I
\&\ V images, which were then de-spiked to remove cosmic rays,
de-rotated to a common reference frame, and temporally averaged using
a Gaussian weighting. The temporal averaging function had a FWHM of
2~frames (1.5~minutes) and preserved the 45-second cadence of the
original data. Finally, the magnetograms were spatially smoothed by
convolving them with a $3\times3$ 2-pixel FWHM Gaussian kernel
$k=\frac{1}{16}\times\left[\begin{array}{ccc} 1 & 2 & 1\\ 2 & 4 &
    2\\ 1 & 2 & 1
\end{array}\right]$ . 

\subsection{Feature and Network Concentration Tracking}
\label{sec:feature-tracking}
For feature tracking, we used the SWAMIS magnetic feature tracking
code\footnote{available at
    http://www.boulder.swri.edu/swamis} \citep{DeForest2007} with
following parameters: thresholds of 18 \&\ 24~G, $x-y-t$ diagonals
disallowed in the detection hysteresis, a per-frame minimum size of
4~pixels, a minimum lifetime of 4~frames, and a minimum total of
8~pixels over each feature's lifetime. We used both the ``downhill''
and the ``clumping'' methods of feature
identification. The ``downhill'' method groups pixels
  into features by finding local maxima (for positive pixels) and
  working downhill until a local minimum is found (and oppositely for
  negative pixels), and is better for finding the small-scale
  structure of the magnetic fields. The ``clumping'' method groups all
  adjacent same-signed pixels into the same feature, and is better for
  finding the large-scale structure of the magnetic fields. For
  example, a large supergranular network concentration will be counted
  as only one feature by the ``clumping'' method, but as many features
  by the ``downhill'' method \citep{DeForest2007,Parnell2009}. The
downhill method identified 112217 features, while the clumping method
found 53910 features throughout the 5.25-hour time span of the
dataset.

\begin{figure*}
\includegraphics[width=1.0\textwidth]{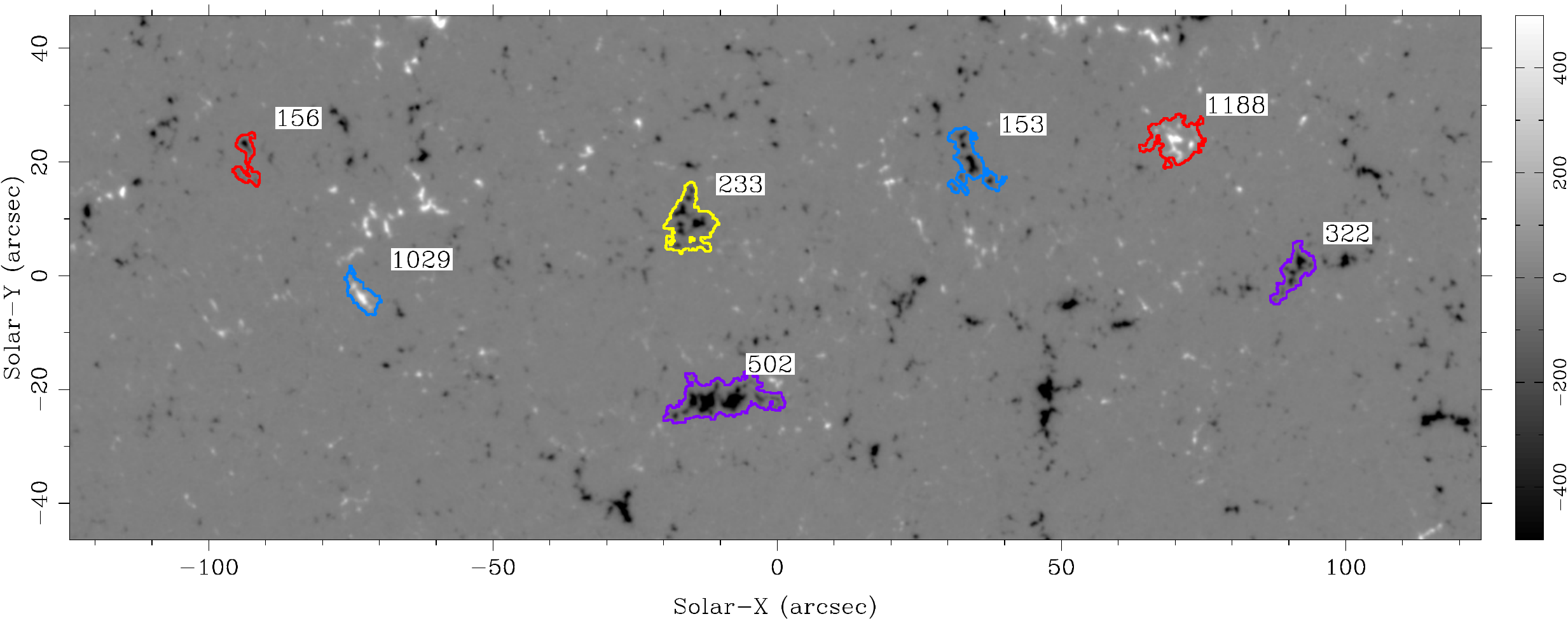}
\caption{The seven network concentrations identified using the SWAMIS
  tracking code described in the text. The background magnetogram is
  the initial frame in the dataset (2007-09-19 at 12:44:44), and the
  colored outlines show the extent of the network concentrations in
  that frame.  The numbers next to each concentration are the feature
  ID numbers SWAMIS assigned while using the ``clumping'' method, and
  are used throughout the paper.}
\label{fig:ID'd-Network-conc}
\end{figure*}

Network concentrations were identified using the clumped feature
tracking data. A feature was designated a network concentration if it
was present for the entire length of the dataset and had a peak flux
density $>500\textrm{ Mx}\textrm{ cm}^{-2}$ in the first frame. We
ignored network concentrations measured within 100 pixels of the edge
of the field of view, so as to remove the complication of edge
effects. We identified seven network concentrations using this method;
they are shown on a single frame in
Figure~\ref{fig:ID'd-Network-conc}. As expected, the identified
network concentrations were separated by distances of 15--30 Mm, which
is approximately the range of supergranule diameters \citep{Meunier2007}.

\subsection{Removing ``Wander''}
\emph{Hinode}'s pointing ``wander'', which translates the frames by many 
pixels on timescales of tens of minutes, posed a problem at the spatial 
scales used in the present study. The wander is not recorded in the 
reported scientific coordinates (arcseconds from disk center) in the 
\emph{Hinode} FITS files, so other means of removing it were required. 

\begin{figure}
\includegraphics[width=0.8\columnwidth]{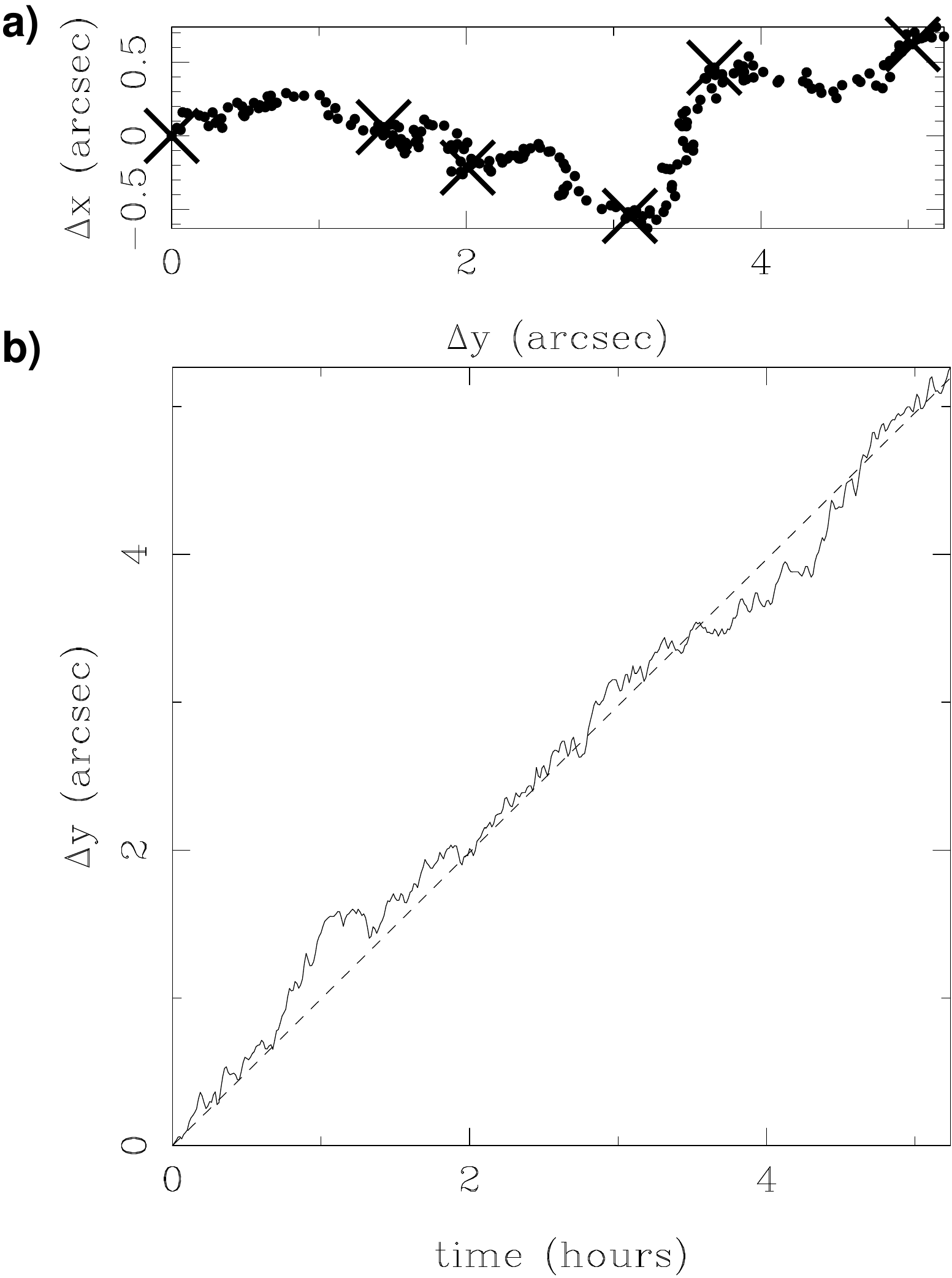}
\caption{\emph{Hinode}'s pointing ``wander'', derived from the median 
horizontal and vertical frame-to-frame change in the position of features.
a) Total wander as a function of time. The beginning of the datasets is at 
(0,0). For clarity, only every other point has been plotted, and note the 
reversed abscissa and ordinate. The large crosses show the wander on  1-hour 
intervals. b) Vertical component of wander plotted with time. The dashed line 
is the first-order least-squares fit that includes the constraint that it 
intersects the origin. Its slope is $0.98"\textrm{ hour}^{-1}$.}
\label{fig:Hinode-pointing-wander}
\end{figure}

We determined a pointing offset between each pair of frames by taking
the median horizontal and vertical displacement across all features
common to those two frames, attributing the net displacement entirely
to residual solar rotation and pointing drift\footnote{\emph{Hinode} was
tracking a region of the solar surface throughout the observation, so
the mean solar rotation in the image sequence was already removed.} (Figure~\ref{fig:Hinode-pointing-wander}a). The 
horizontal drift varied between $-0.63"$ and $0.75"$, but we found a more 
steady positive vertical drift. A linear least-squares fit to the vertical 
drift, requiring that the line intersect the origin, reveals a 
$0.98"\textrm{ hour}^{-1}$ drift northward (Figure~\ref{fig:Hinode-pointing-wander}b).

The drift was derived from the clumping tracking method; the drift
derived from downhill tracking has the same overall shape but is
systematically smaller by about 5\%. We attribute this to greater
``noise'' in the clumping signal, because the flux-weighted
center-of-gravity of the larger clumped features is more affected by
merging and fragmenting of flux.  Nevertheless, we use the drift
derived from the clumping dataset, since that method was used for
network concentration identification (Section~\ref{sec:feature-tracking}).

\section{Statistical Analysis \& Results}
\label{sec:results}
Our objective was to identify whether there is a tendency for new
features to cluster or avoid regions near to or away from network
concentrations. This required a statistical evaluation of the location
and polarity of new features compared with that of the
concentration. Computer code used to
  conduct the statistical analysis and produce the figures can be
  obtained by contacting the lead author.

\subsection{The Position of New Features Around Network Concentrations}
\label{sub:Position-New-Features}
\begin{figure}
\includegraphics[width=0.9\columnwidth]{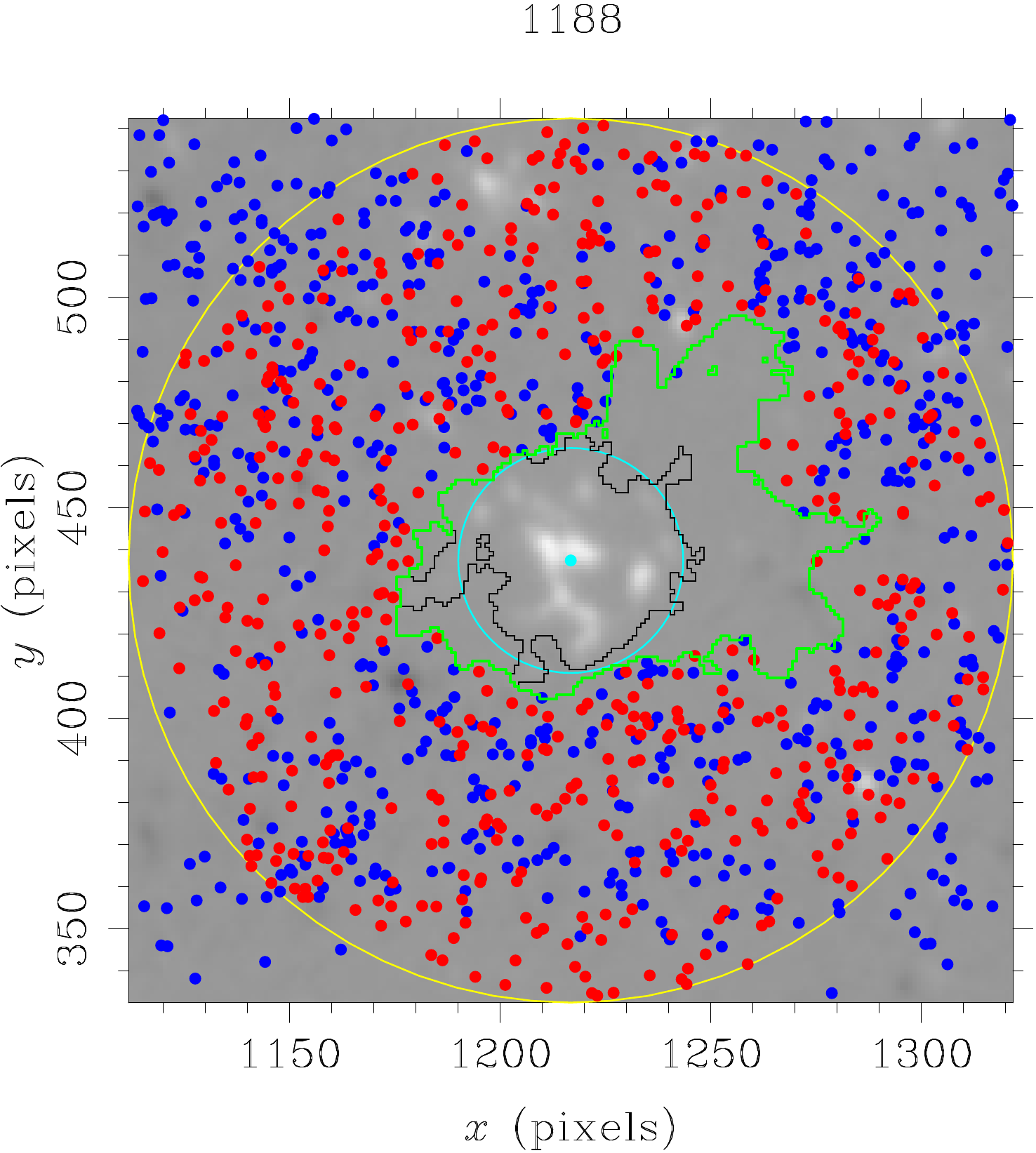}
\caption{A selected $212\times 212$ pixel area on the first frame 
(2007-09-19 at 12:44:44UT, chosen purely for convenience) in our dataset 
showing the network concentration numbered 1188 in 
Figure~\ref{fig:ID'd-Network-conc}. Background: magnetogram from the 
selected area in this first frame. Cyan dot and circle: center-of-flux 
and initial circularized area of the concentration. Black outline: initial 
area of the concentration. Green outline: maximum spatial extent of the 
concentration. Blue dots: locations of new features (not born by Fragmentation 
or Error) during the entire sequence. Red dots: random locations, the generation 
of which is described in Section~\ref{sec:random}. Note that the blue dots 
indicate the locations of new features throughout the entire dataset 
and do not contain information on the timing of the feature's appearance.}
\label{fig:Net-Conc-surroundings}
\end{figure}

While the clumping method allows for easier identification of the network 
concentrations, more precise spatial locations are obtained with the 
downhill method since the same ``clumped'' feature is broken into 
several smaller features. Thus, for every frame, we found those downhill 
features that were at the location of the network concentration identified 
in the clumped data. Throughout the dataset, each network concentration 
changed in size and shape (this is explored further in 
Section~\ref{sub:NetworkConcEvolution}), and so together these downhill 
features describe the network concentration's maximum spatial extent (green
outline in Figure~\ref{fig:Net-Conc-surroundings}), adjusted to accommodate 
for the \emph{Hinode} pointing wander.

We identified all small-scale features within 100 pixels (11.6~Mm) of
the network concentrations' initial flux-weighted center-of-gravity
(the cyan dot in Figure~\ref{fig:Net-Conc-surroundings}). We excluded
those features that were present at the beginning of the dataset,
along with those that were created via the ``Fragmentation'' or
``Error'' categories \citep[see][for descriptions of these categories
  of feature creation]{Lamb2008,Lamb2010}.  The locations of the
selected features are indicated by the blue dots in
Figure~\ref{fig:Net-Conc-surroundings}, superimposed on top of a
section of the first frame in the dataset. This shows every new
feature location regardless of the time at which it was created, so
there is little-to-no temporal relationship between the underlying
magnetogram in the figure and the blue dots. The blue dot locations
also accommodate for wander.

\subsection{Monte Carlo Simulations}
\label{sec:random}
\begin{figure}
\includegraphics[width=1.0\columnwidth]{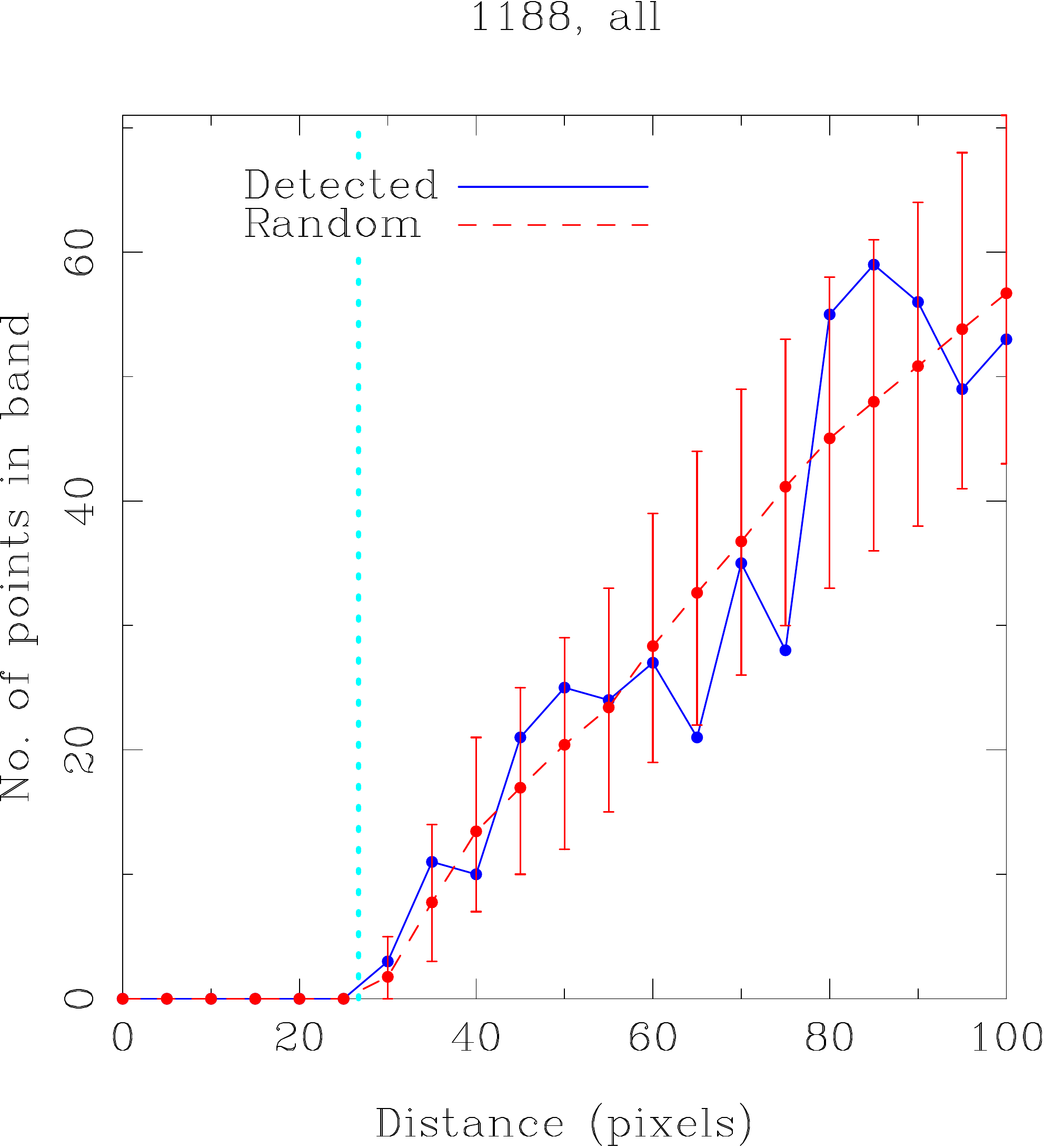}
\caption{For network concentration 1188 (the same as that shown in 
Figure~\ref{fig:Net-Conc-surroundings}), the number
of points in a 5-pixel wide annulus centered on the cyan dot. The
vertical cyan dotted line corresponds to the radius of the cyan circle
in Figure~\ref{fig:Net-Conc-surroundings}. The blue points correspond
to the blue dots (detected features) in Figure~\ref{fig:Net-Conc-surroundings}.
The red points are the mean of $10^{4}$ iterations of the random
placement, the error bars correspond to the 95th percentile around
the mean.}
\label{fig:Net-Conc-Annuli}
\end{figure}

In order to test whether there is a statistical tendency for the number of new features to be enhanced or suppressed in proximity to a network concentration, we measured 
their birth locations against that of a generated set of 
pseudo-random points. For each network concentration, we assigned points using a Monte Carlo 
procedure with $10^4$ iterations within the 200-pixel diameter circle (the 
yellow circle in Figure~\ref{fig:Net-Conc-surroundings}). These are hereafter 
referred to as random points, are shown as red dots in 
Figure~\ref{fig:Net-Conc-surroundings}, and were assigned according to the 
following criteria: 
\begin{enumerate}
\item The number of random points (red dots in Figure~\ref{fig:Net-Conc-surroundings}) 
was equal to the number of detected features (blue dots) inside the circle;
\item Random points were not placed within the maximum spatial extent of 
the network concentration (outlined in green in Figure~\ref{fig:Net-Conc-surroundings}).
\end{enumerate}
The number of points in a 5-pixel-wide annulus, centered on the center-of-gravity 
of the network concentration, and plotted against the radius of the center of 
the annulus, is shown in Figure~\ref{fig:Net-Conc-Annuli}. The blue points 
are the numbers of blue dots (new features) in Figure~\ref{fig:Net-Conc-surroundings} 
within the annulus, and the vertical dotted line is the radius of the circularized 
initial area of the concentration (the cyan circle in 
Figure~\ref{fig:Net-Conc-surroundings}). The red points are the mean of the 
$10^4$ iterations of the random placement, and the error bars correspond to 
the 95th percentile around the mean.

Statistically, this graph can be interpreted as follows: If the
placement of the new features lies within the uncertainty bounds of
the random distribution, then they too are most likely randomly
distributed, i.e., there is no statistical bias towards new features
forming at that particular distance from the network concentration.

Plots similar to Figures~\ref{fig:Net-Conc-surroundings} 
and~\ref{fig:Net-Conc-Annuli}, but for the remaining six network 
concentrations, are shown in Figures~\ref{fig:New-features-(short)} and \ref{fig:New-features-(short)-2}. The three 
plots correspond to three different annular widths, of 5, 10 and 15~pixels.
For the 15-pixel wide annuli, the radius of the region of interest was 
expanded to 105~pixels.

\rotFPtop=0pt

\begin{sidewaysfigure*}

       \includegraphics[width=0.2\textwidth]{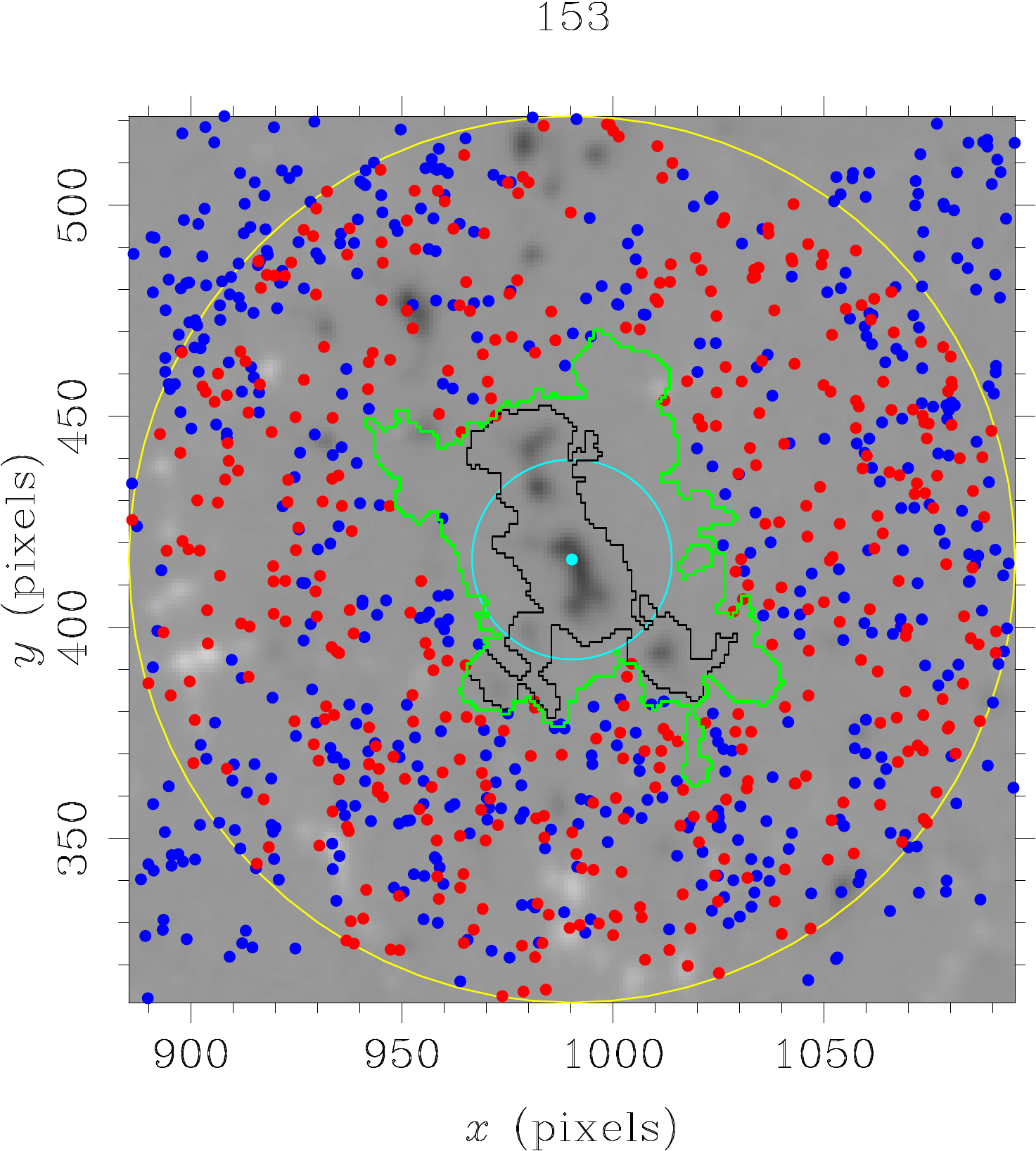}
\hfill \includegraphics[width=0.2\textwidth]{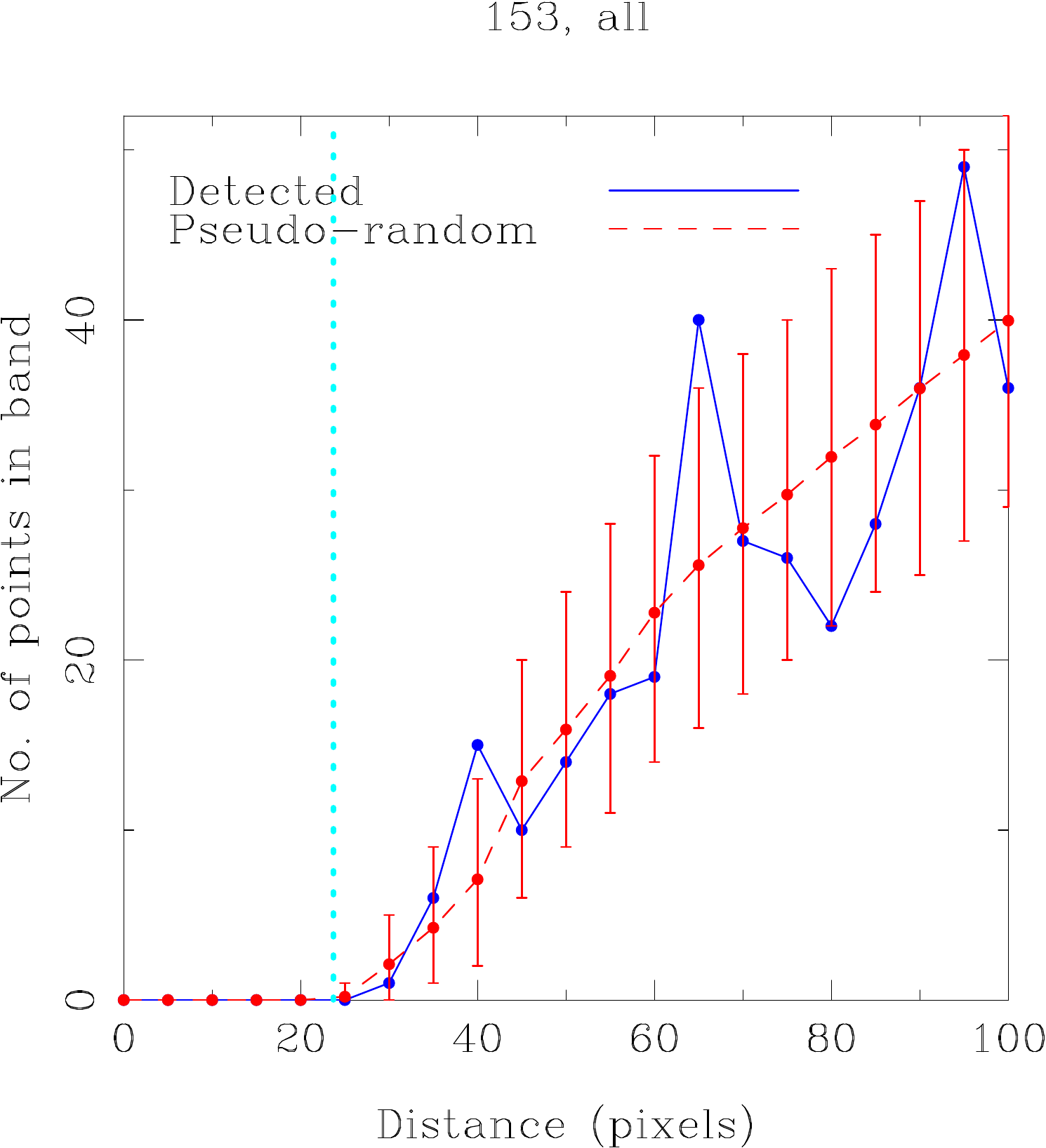}
\hfill \includegraphics[width=0.2\textwidth]{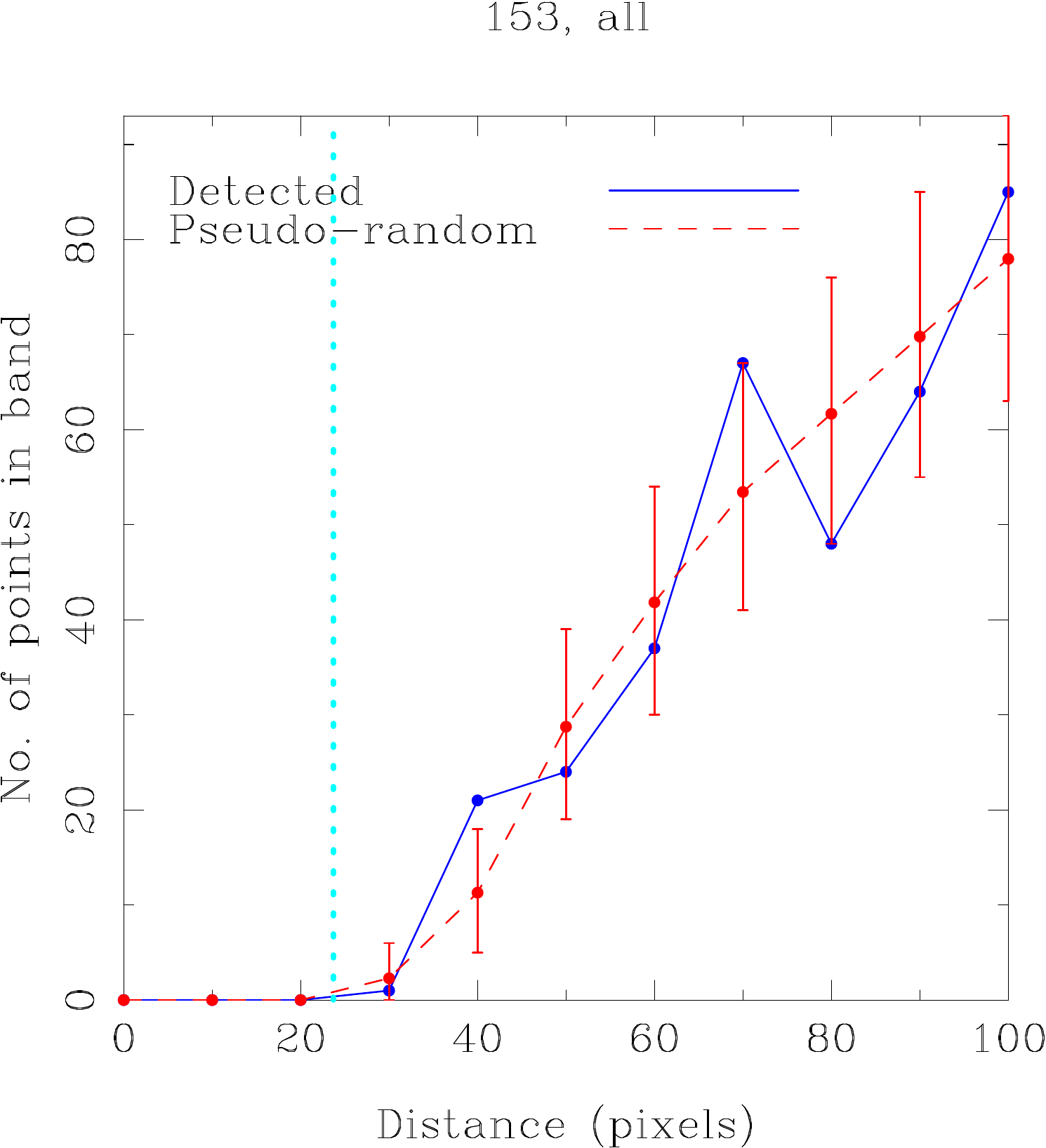}
\hfill \includegraphics[width=0.2\textwidth]{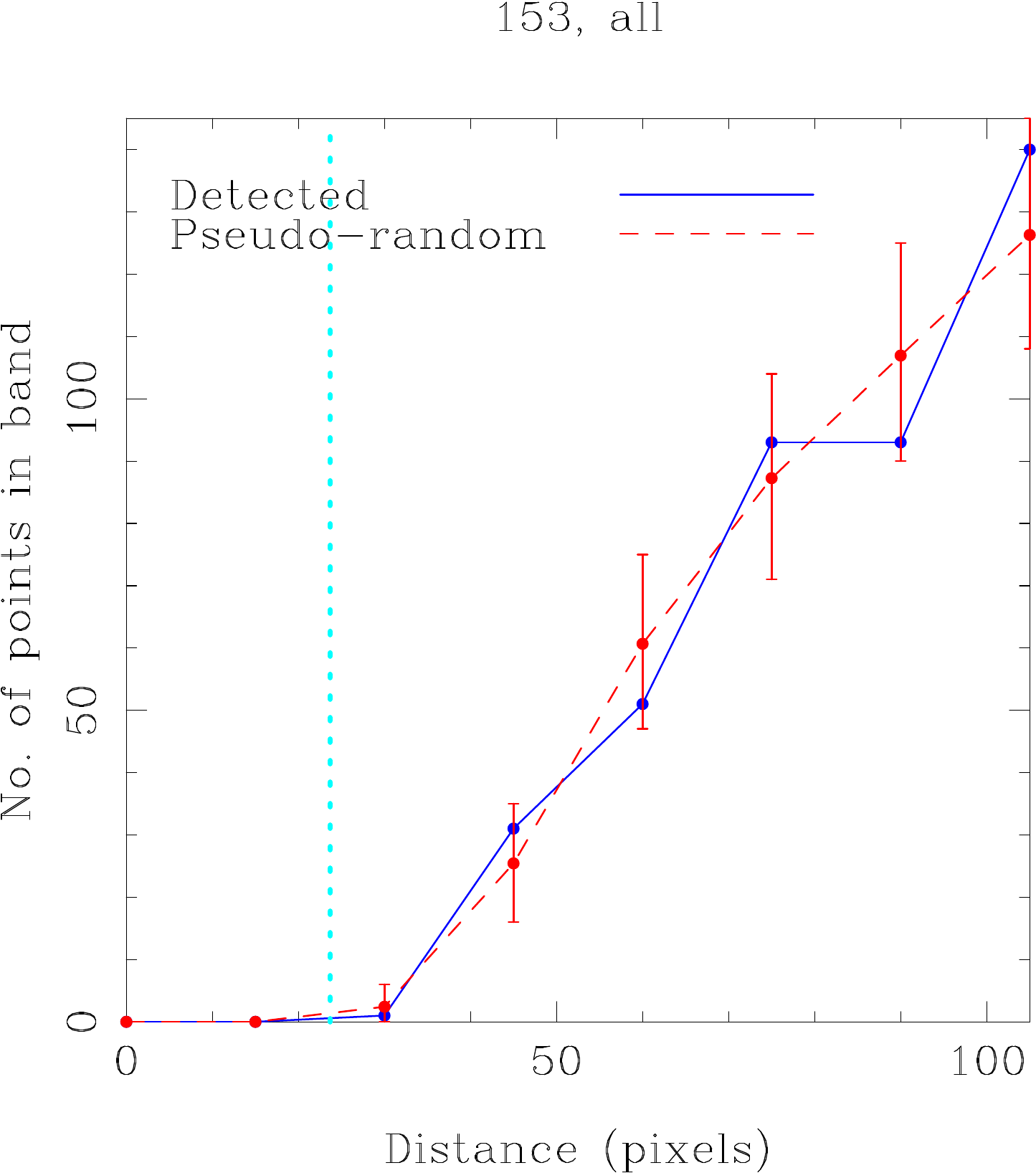}
\medskip
\hfill \includegraphics[width=0.2\textwidth]{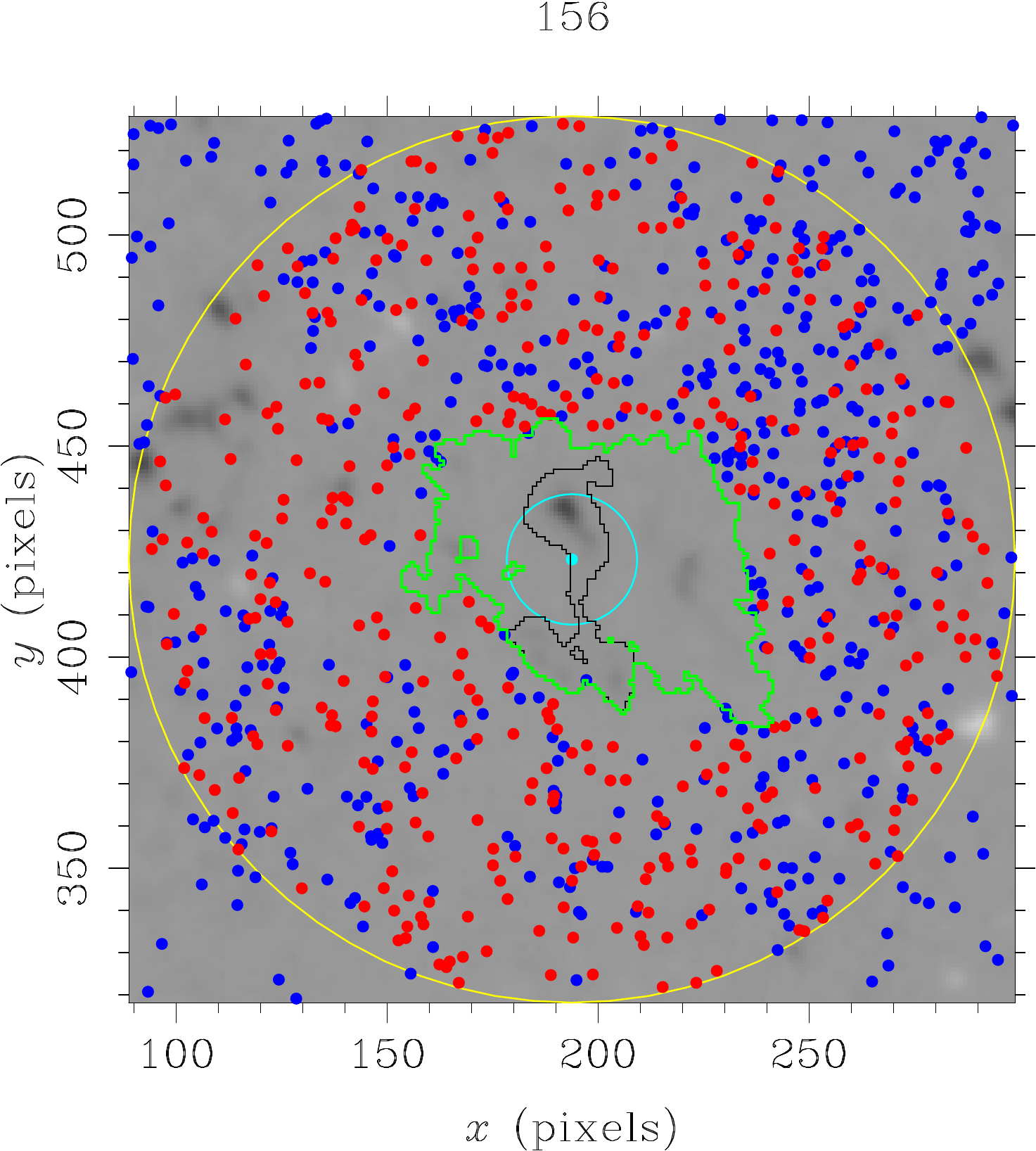}
\hfill \includegraphics[width=0.2\textwidth]{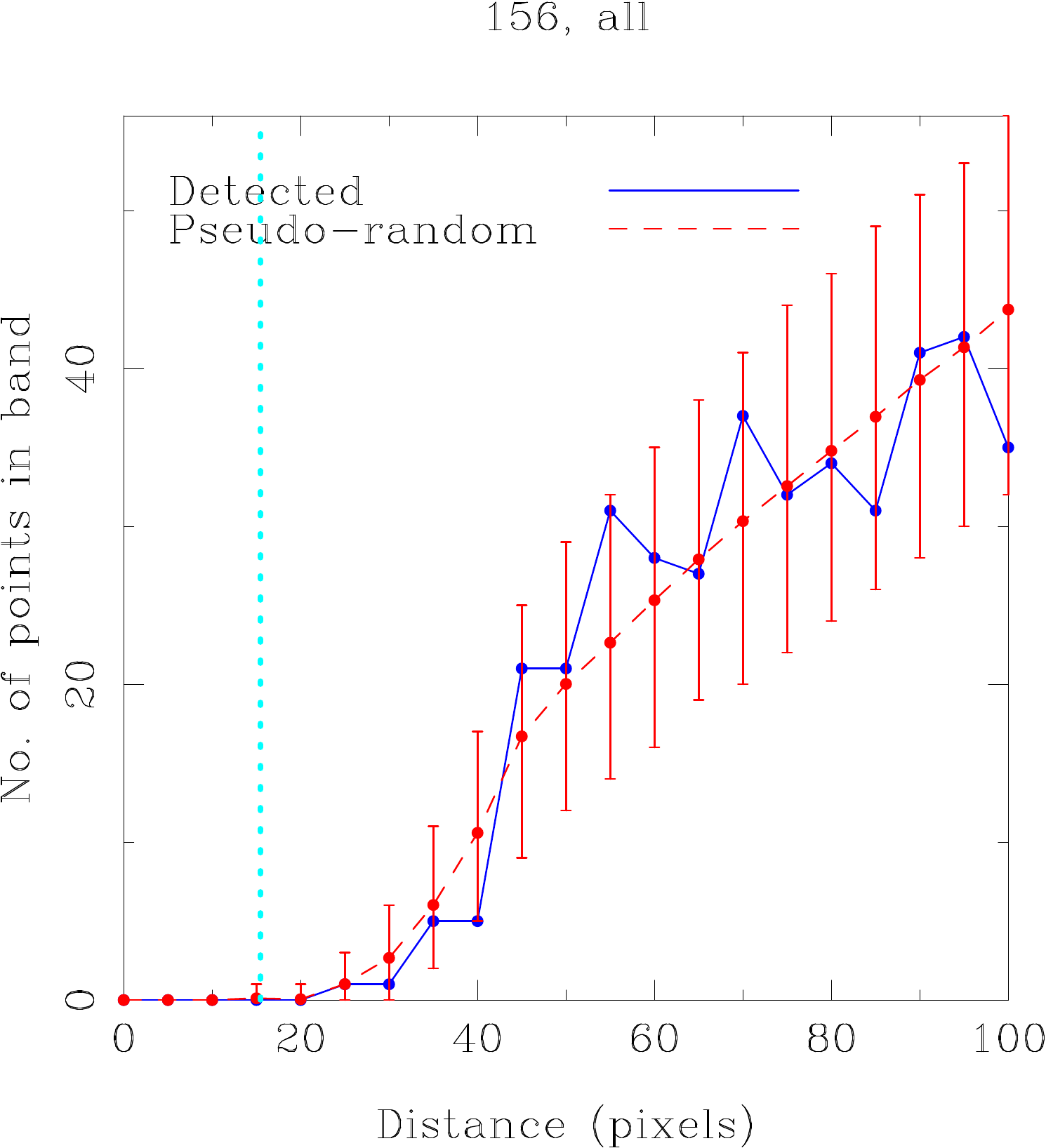}
\hfill \includegraphics[width=0.2\textwidth]{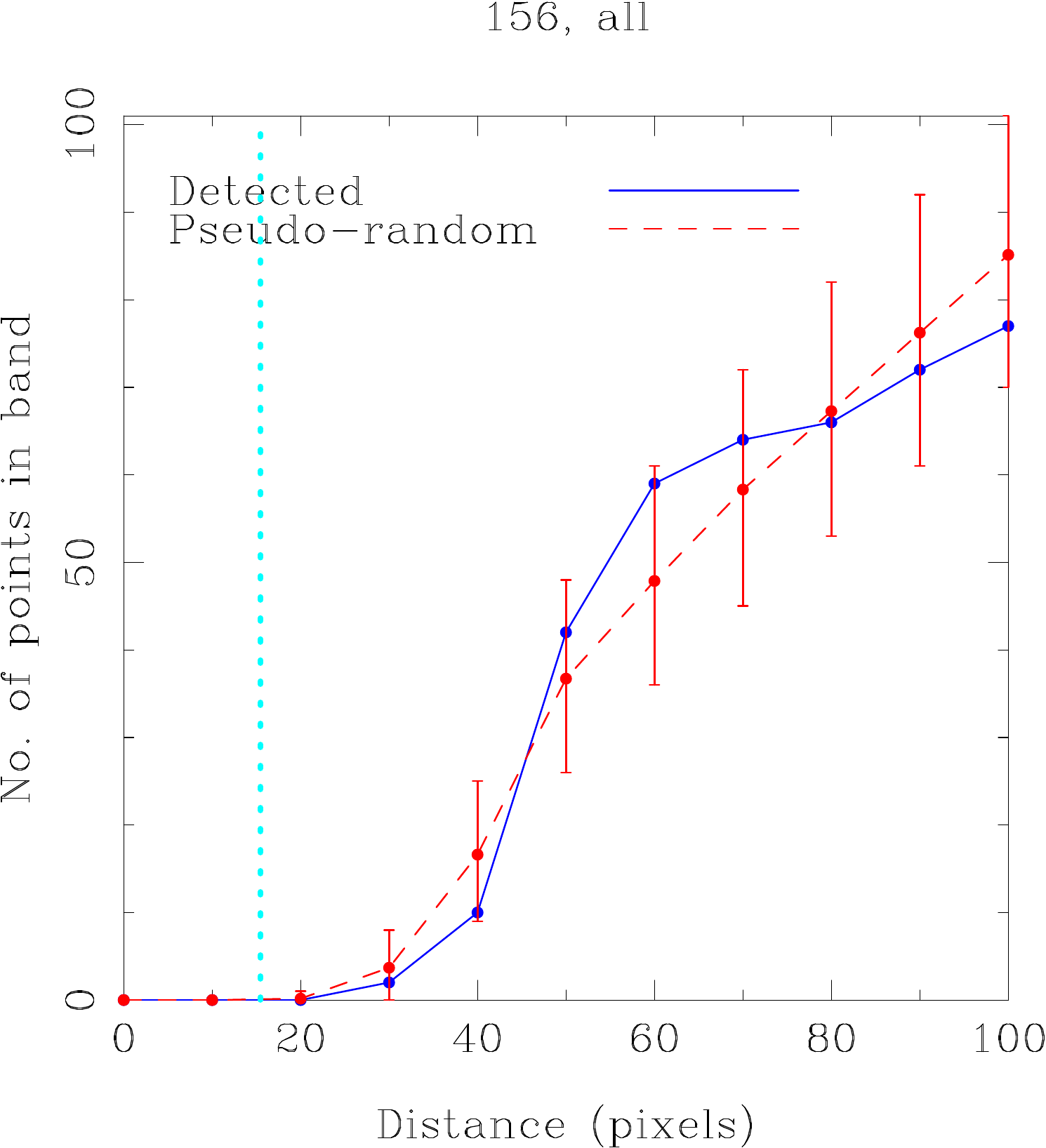}
\hfill \includegraphics[width=0.2\textwidth]{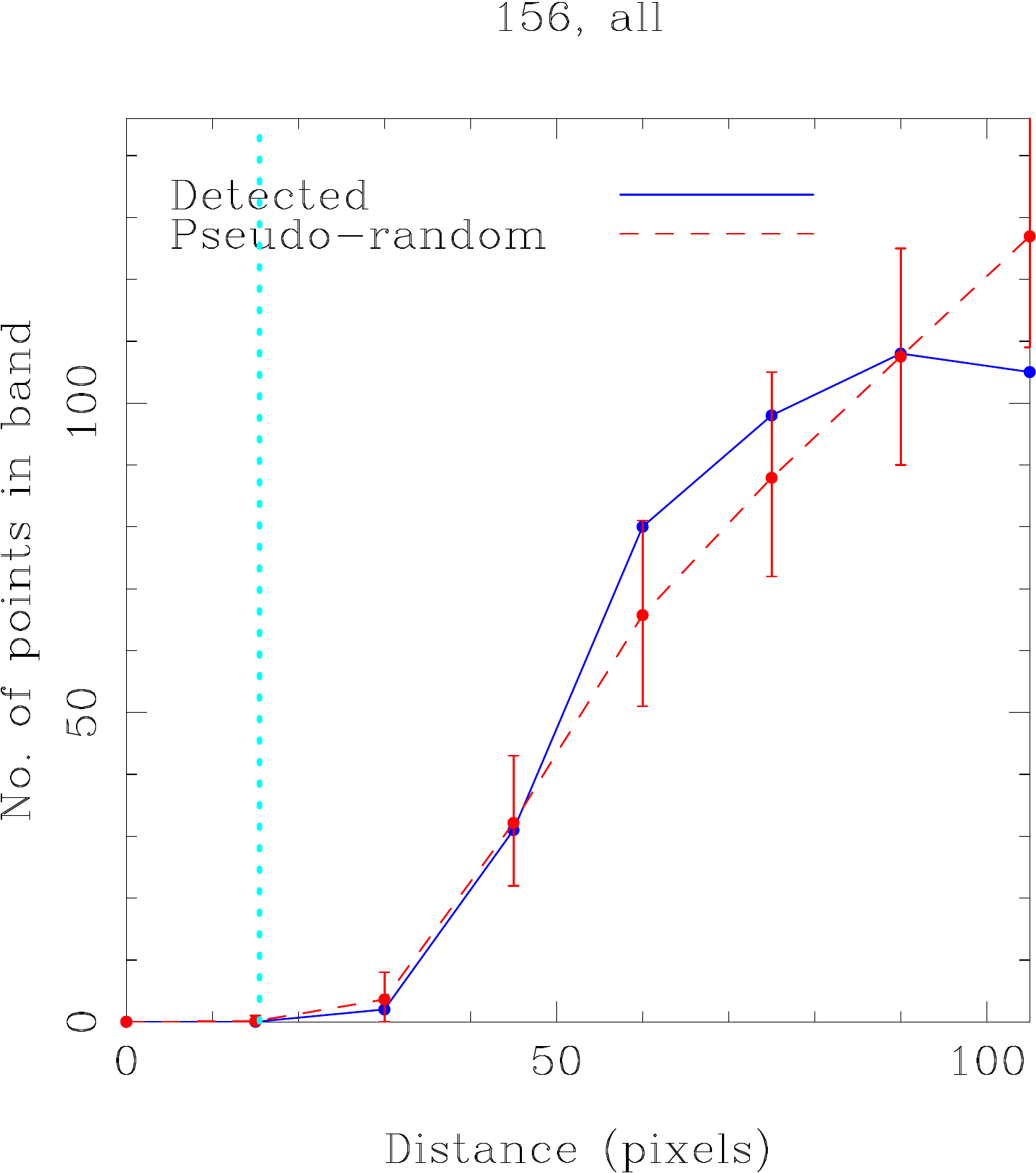}
\medskip
\hfill \includegraphics[width=0.2\textwidth]{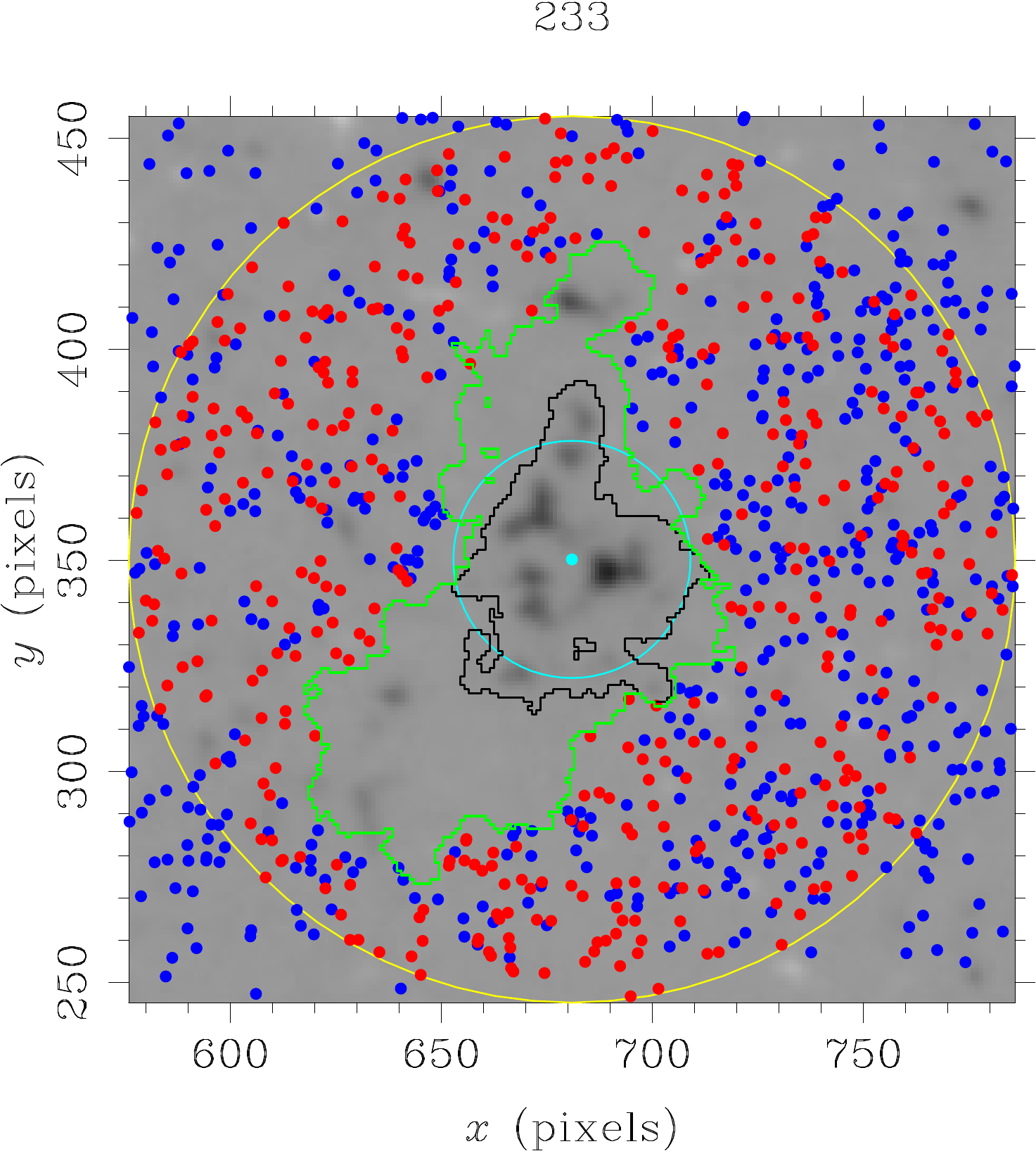}
\hfill \includegraphics[width=0.2\textwidth]{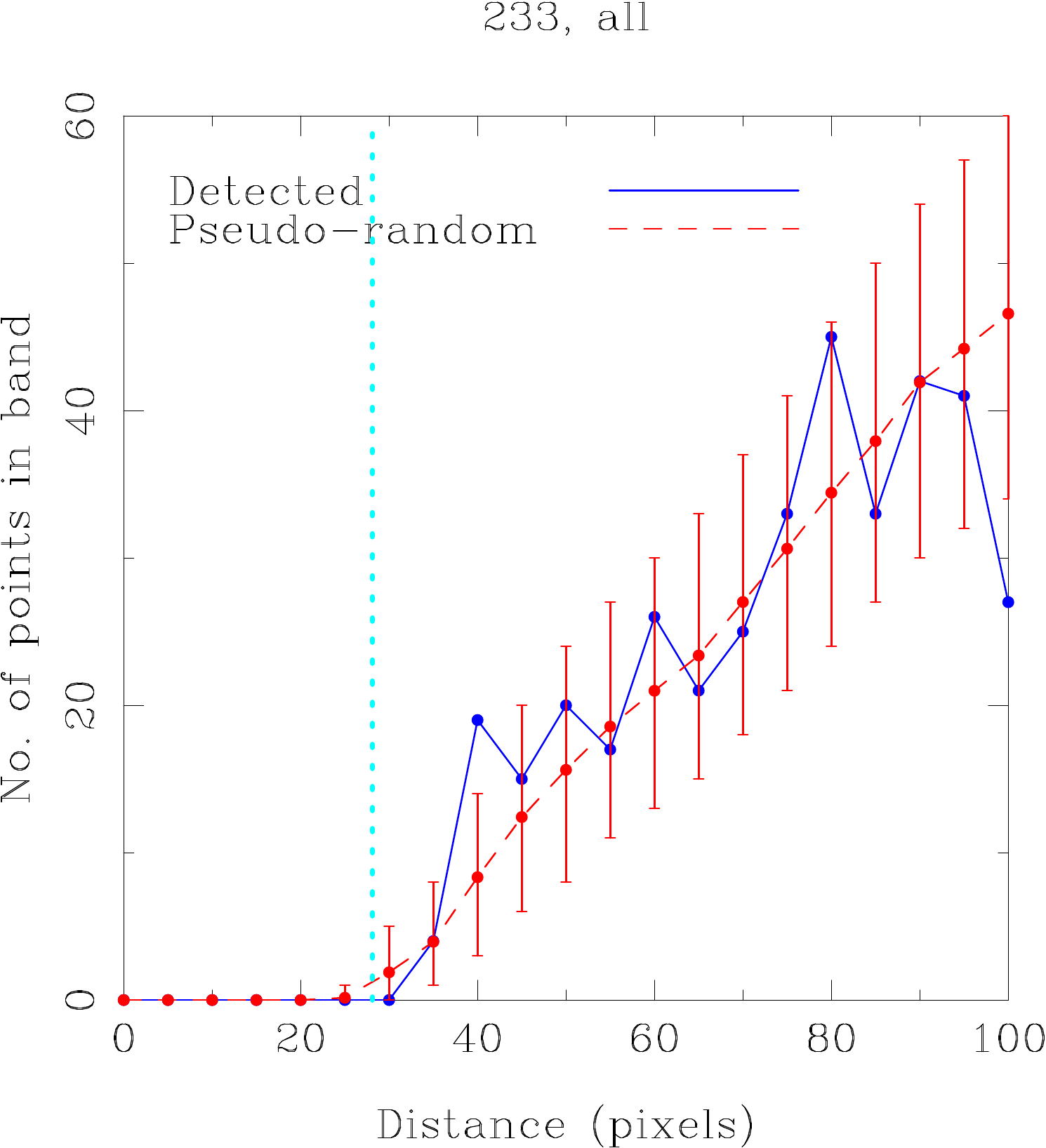}
\hfill \includegraphics[width=0.2\textwidth]{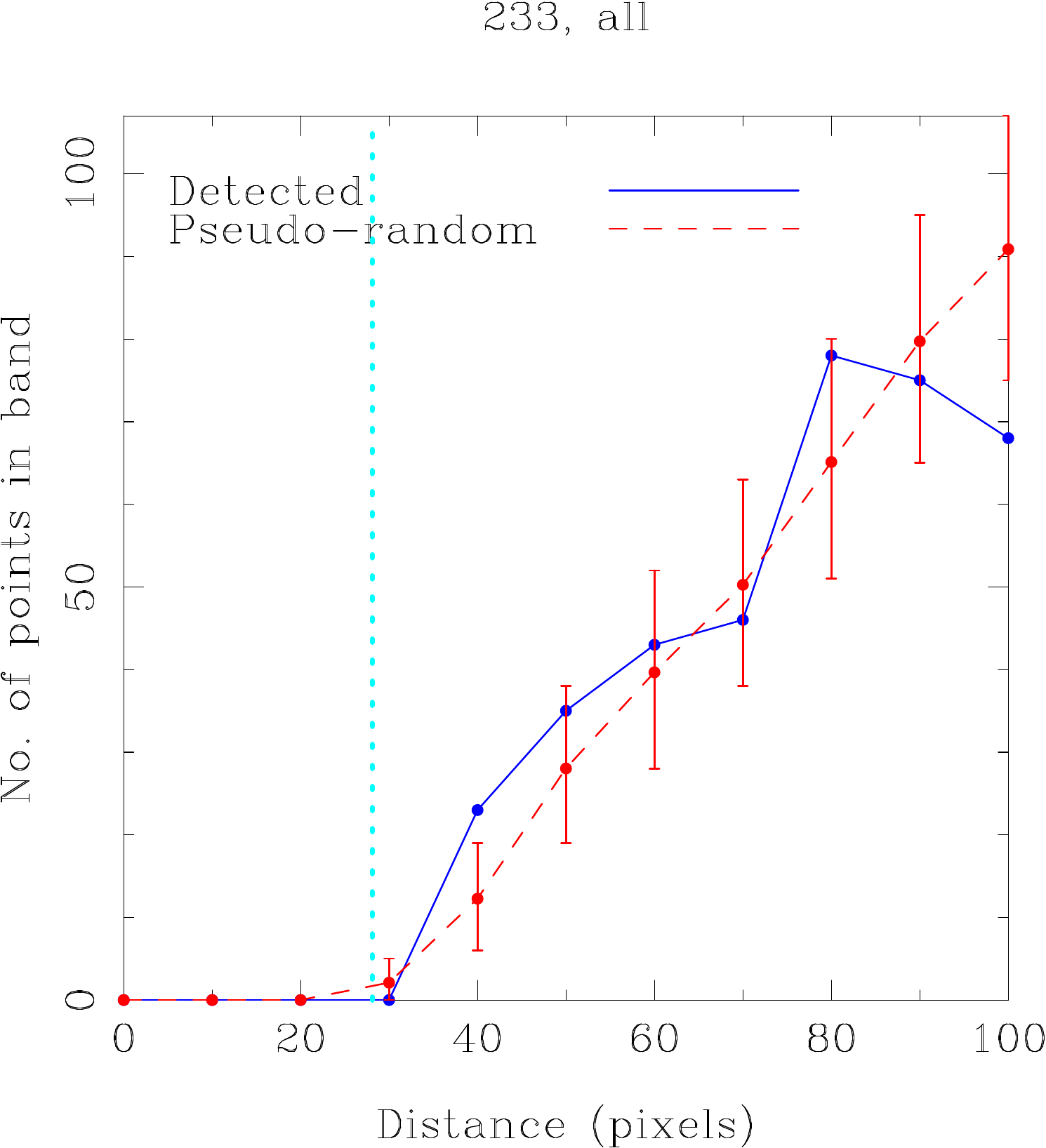}
\hfill \includegraphics[width=0.2\textwidth]{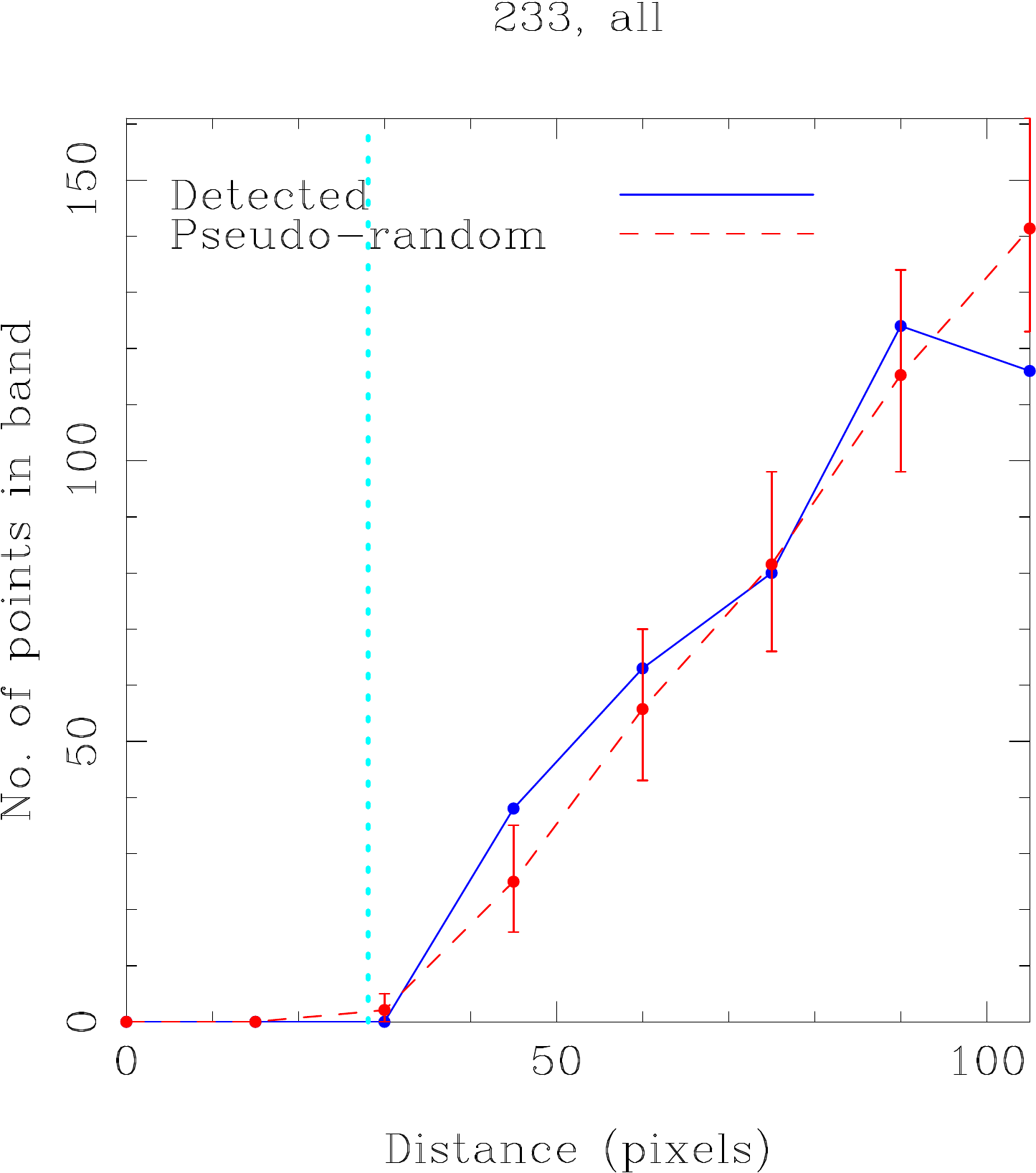}

\caption{Locations of new features around the remaining six network
concentrations, with plots for three different annular widths of 
5, 10 and 15~pixels.}
\label{fig:New-features-(short)}
\end{sidewaysfigure*}

\rotFPbot=0pt
\newpage
\rotFPtop=250pt

\begin{sidewaysfigure*}
 \includegraphics[width=0.2\textwidth]{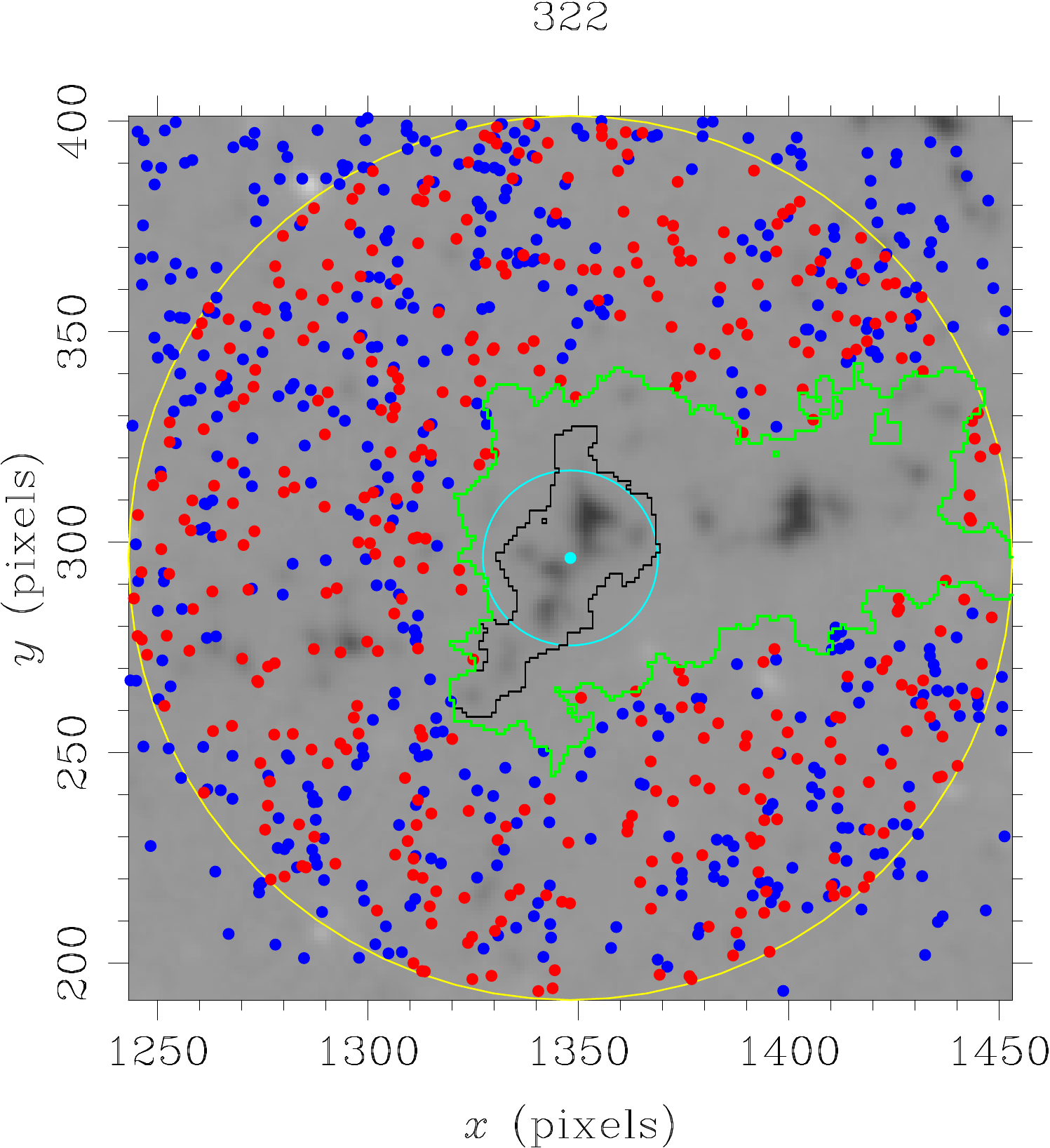}
 \includegraphics[width=0.2\textwidth]{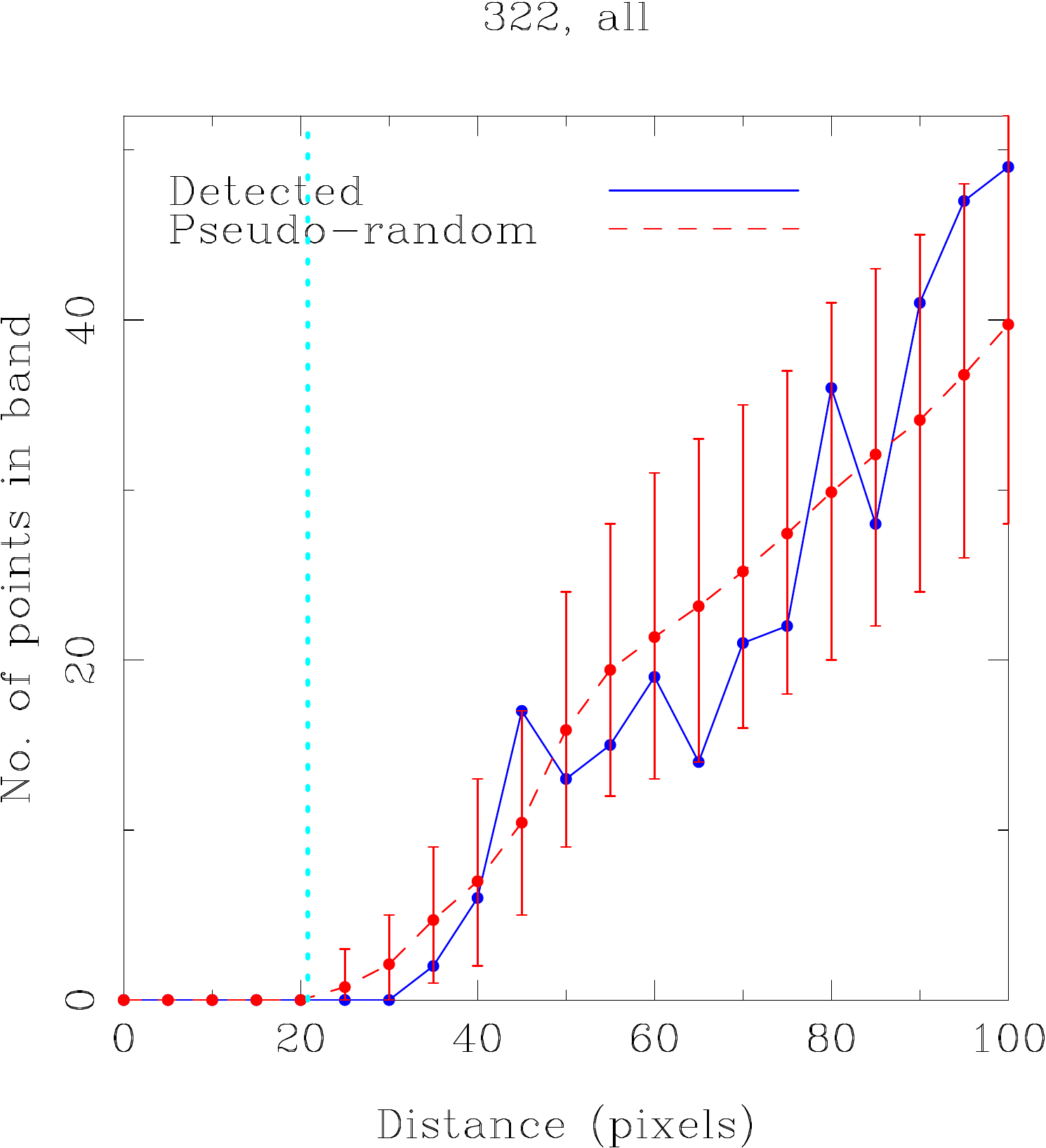}
 \includegraphics[width=0.2\textwidth]{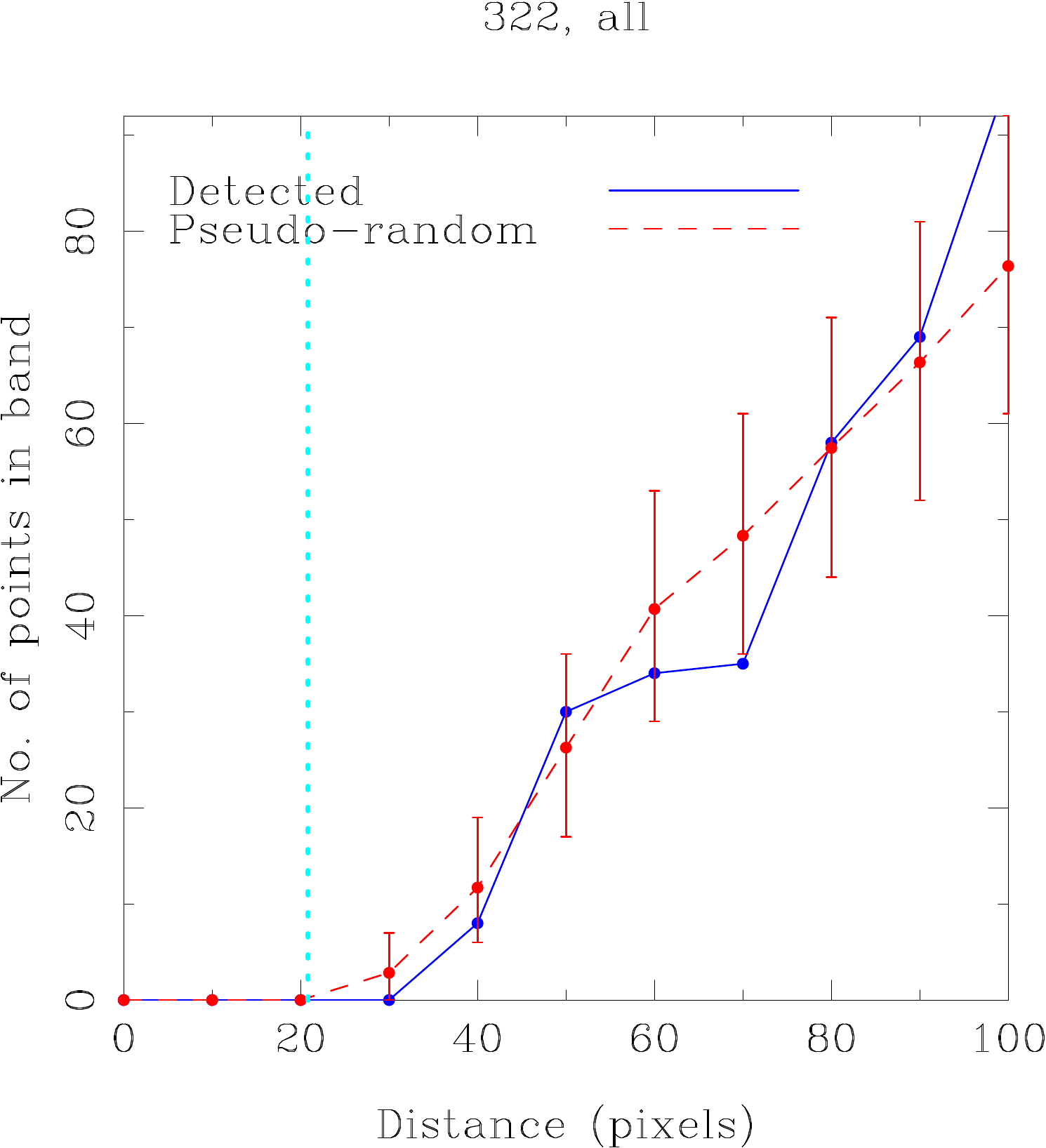}
 \includegraphics[width=0.2\textwidth]{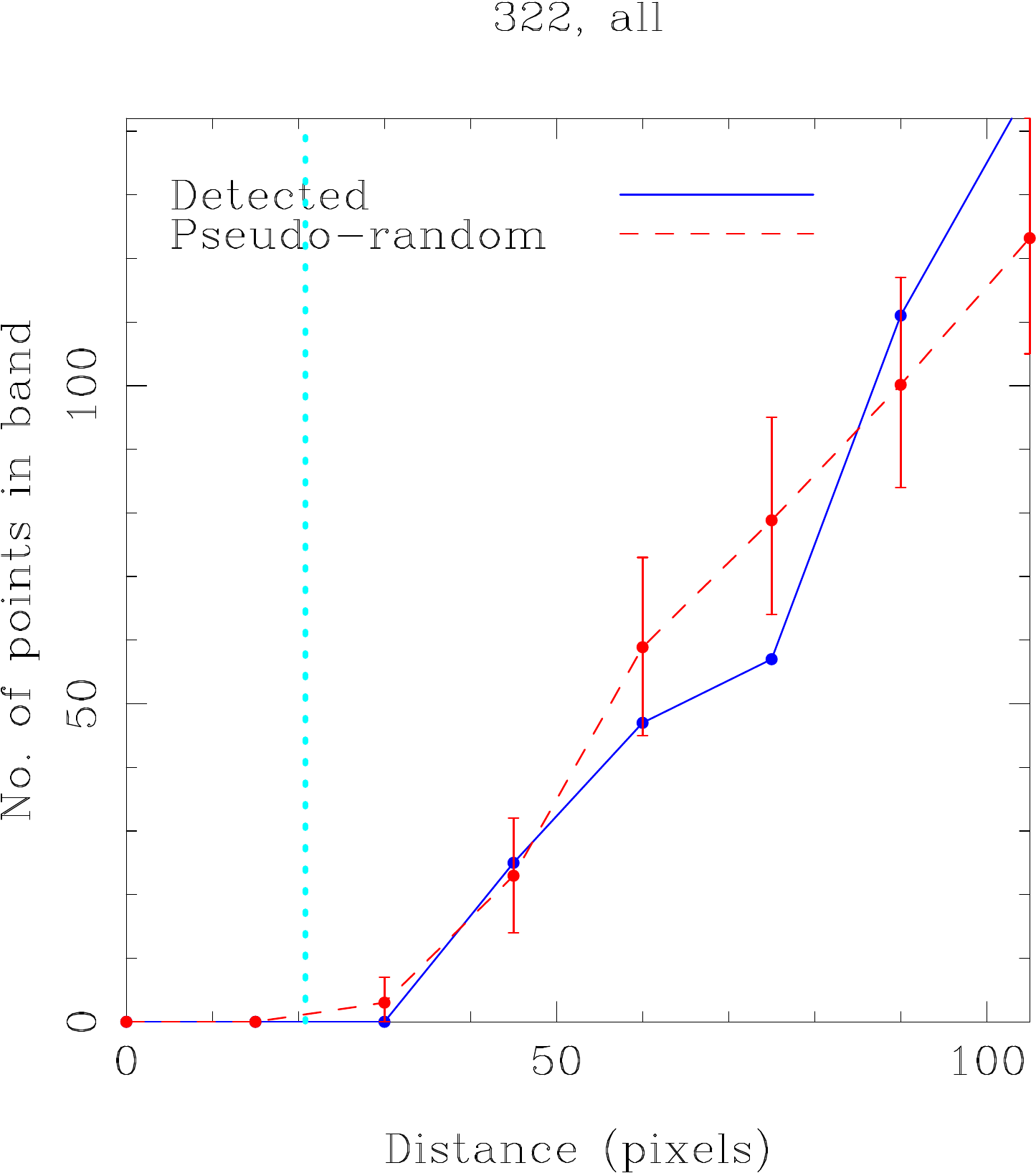}
 \includegraphics[width=0.2\textwidth]{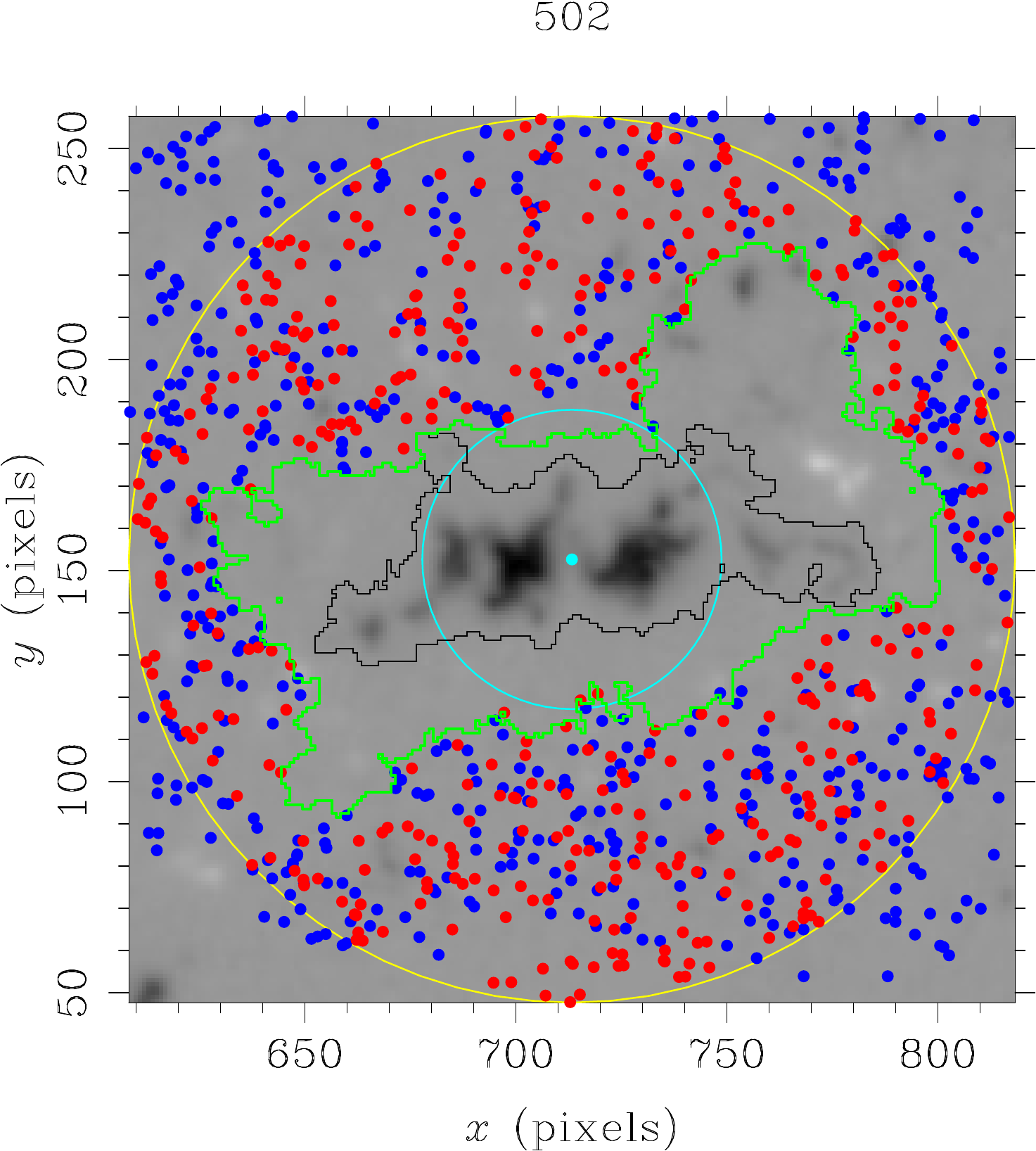}
 \includegraphics[width=0.2\textwidth]{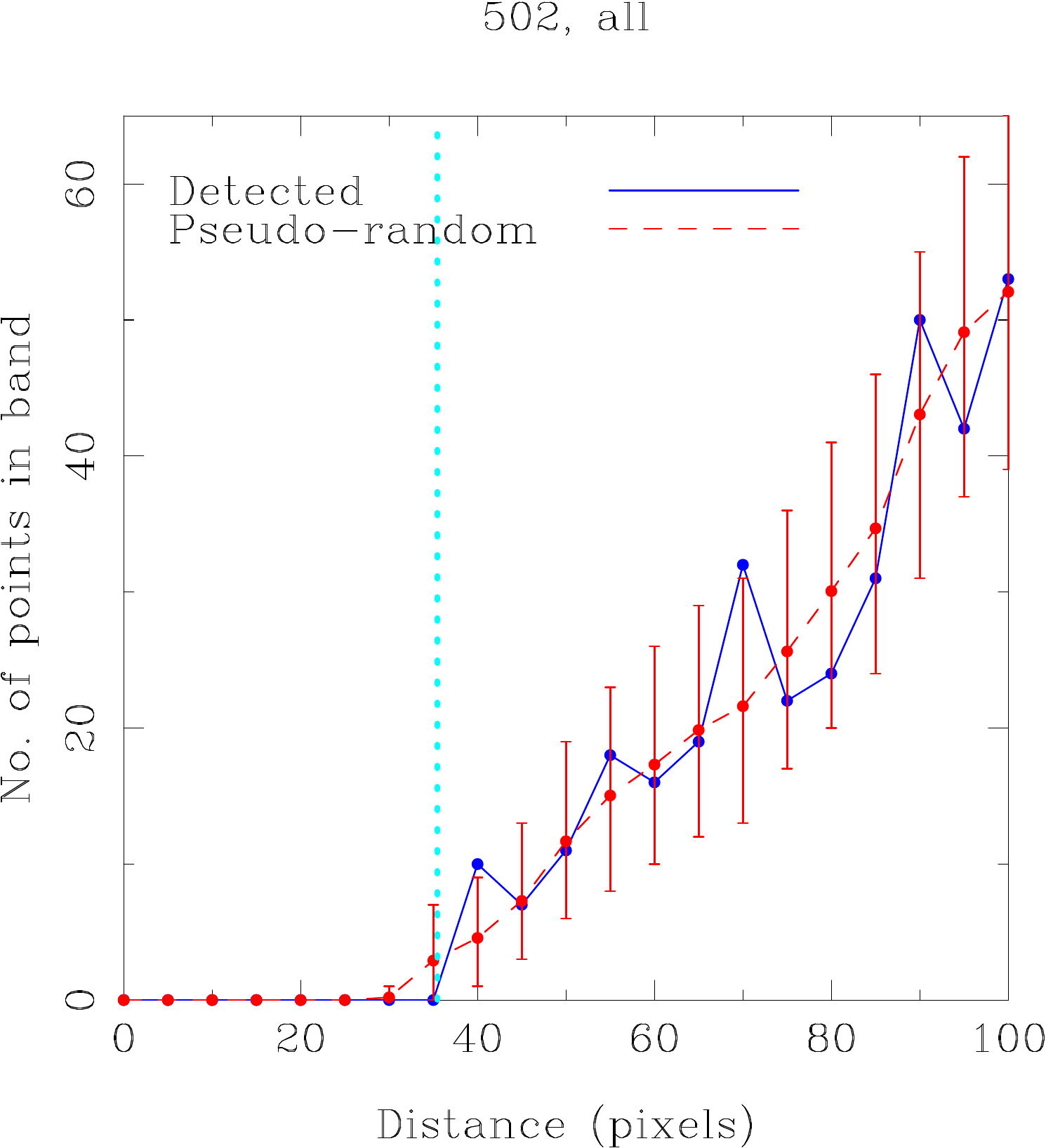}
 \includegraphics[width=0.2\textwidth]{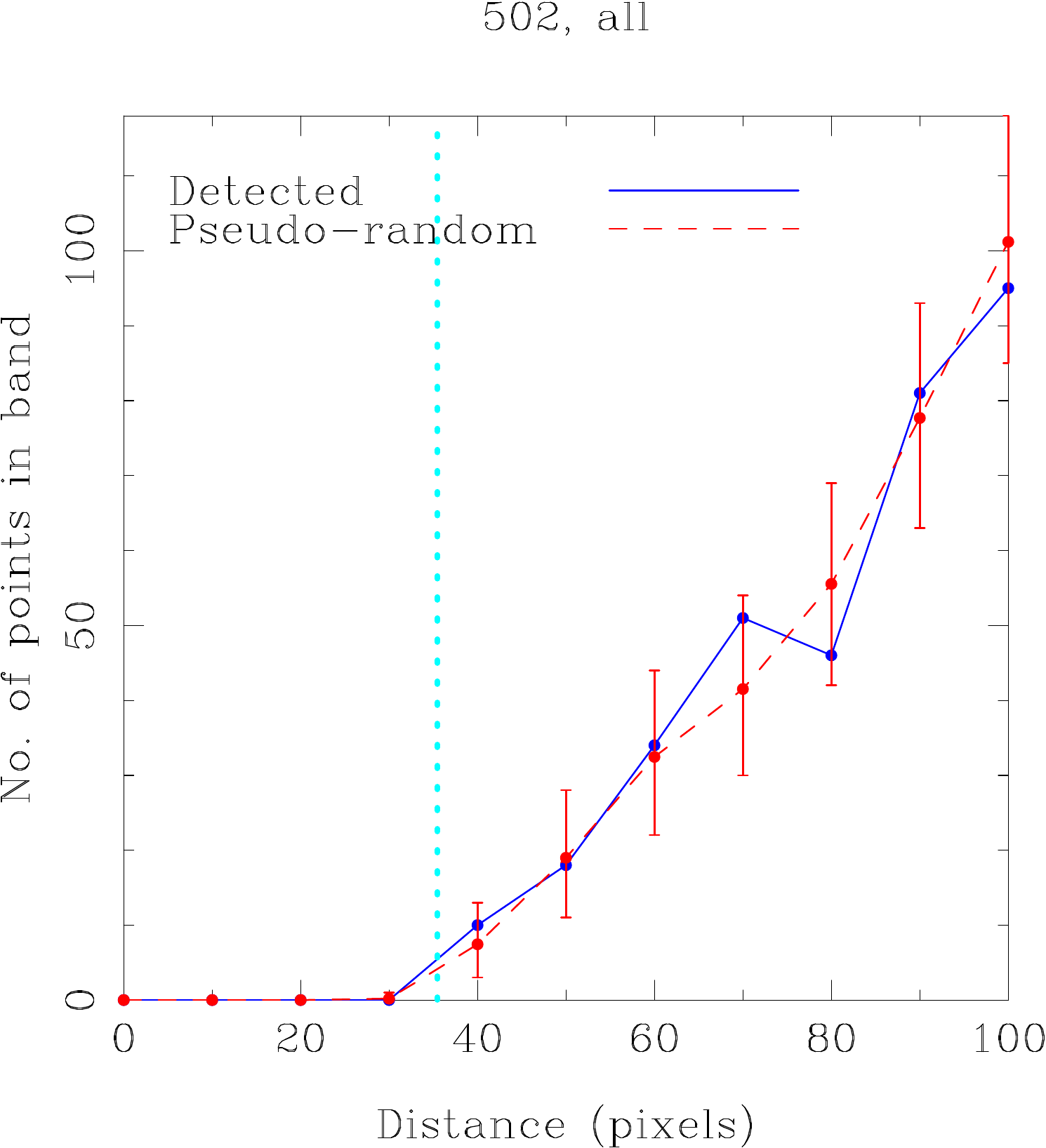}
 \includegraphics[width=0.2\textwidth]{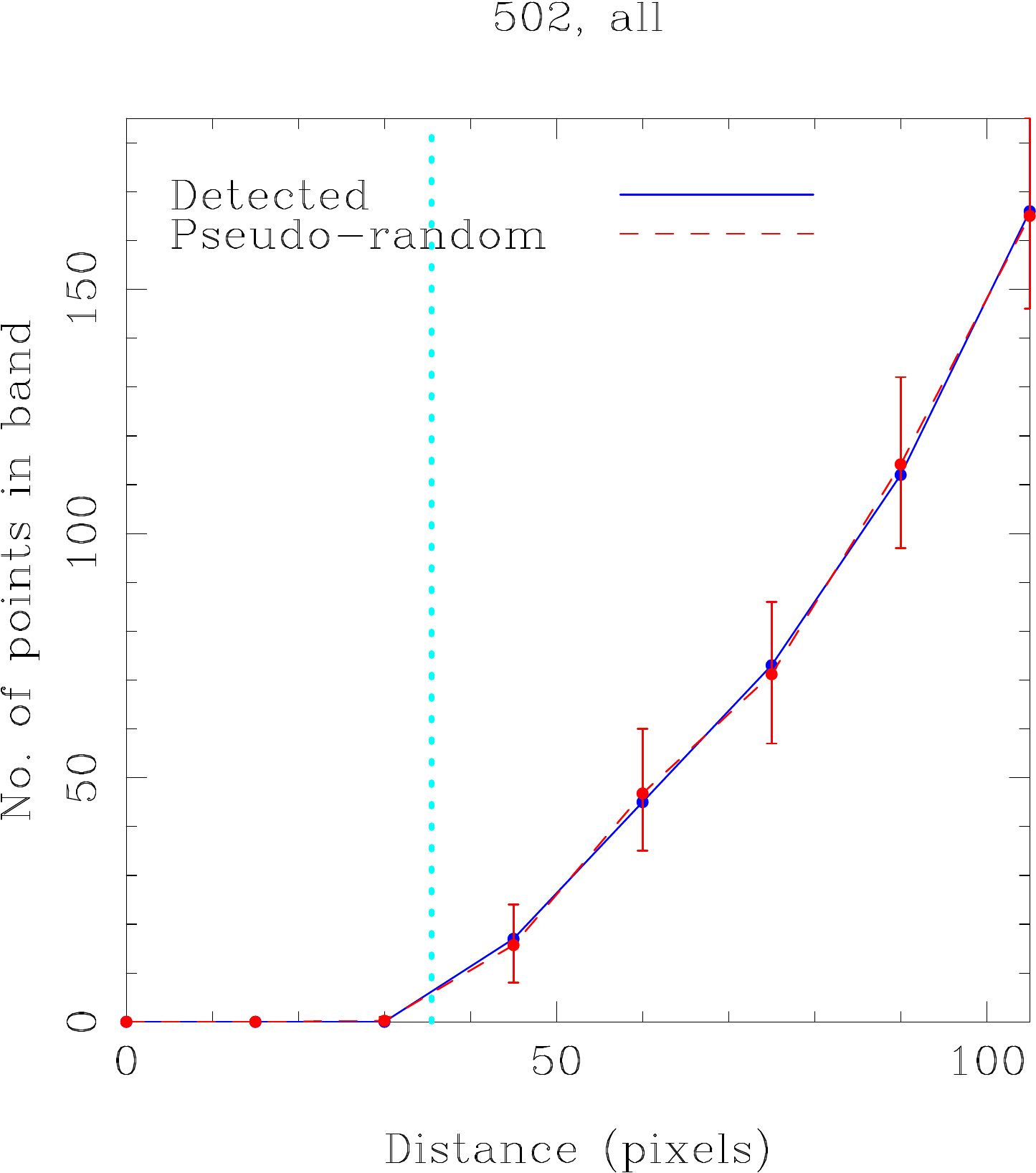}
\hfill \includegraphics[width=0.2\textwidth]{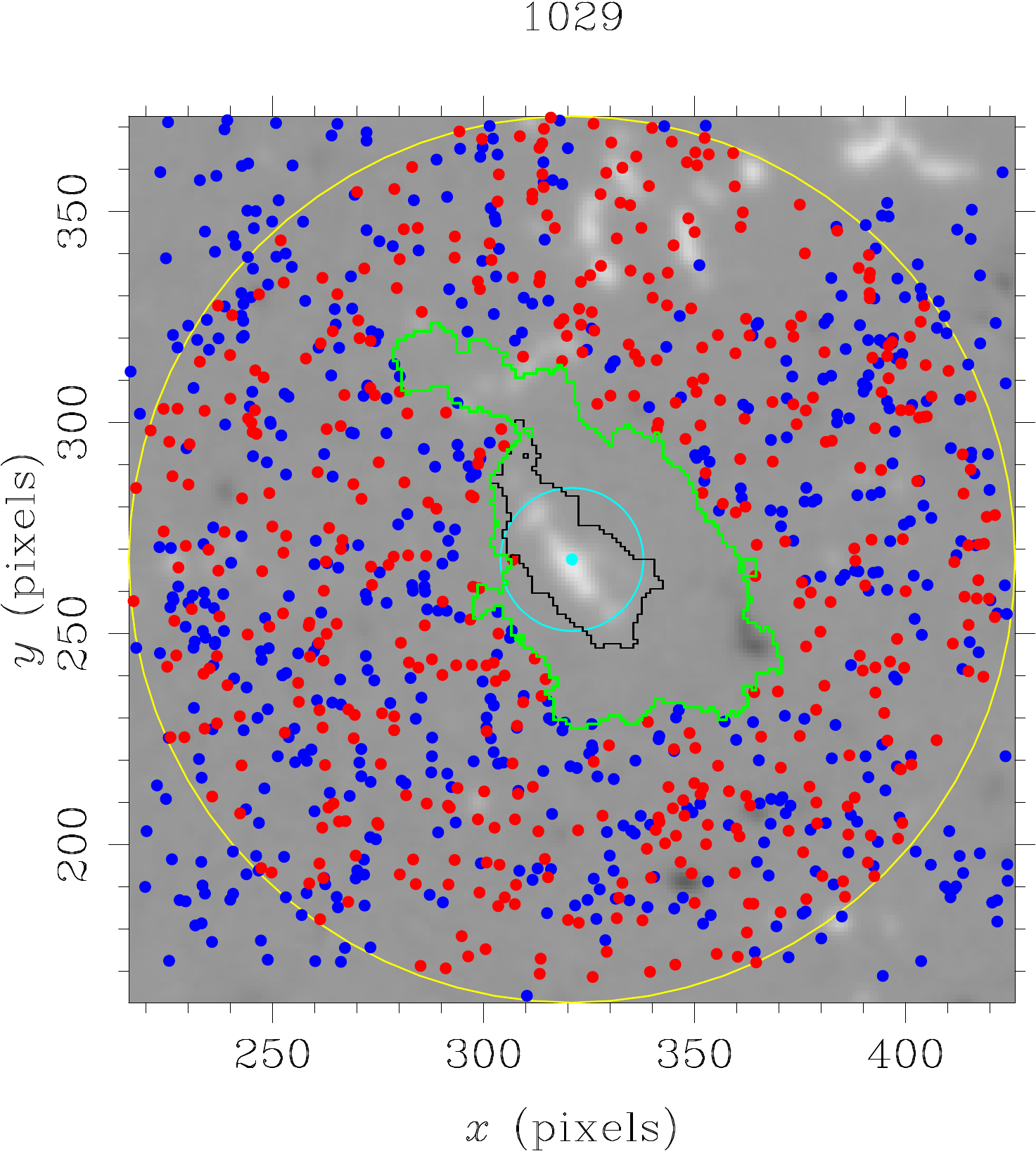}
\hfill \includegraphics[width=0.2\textwidth]{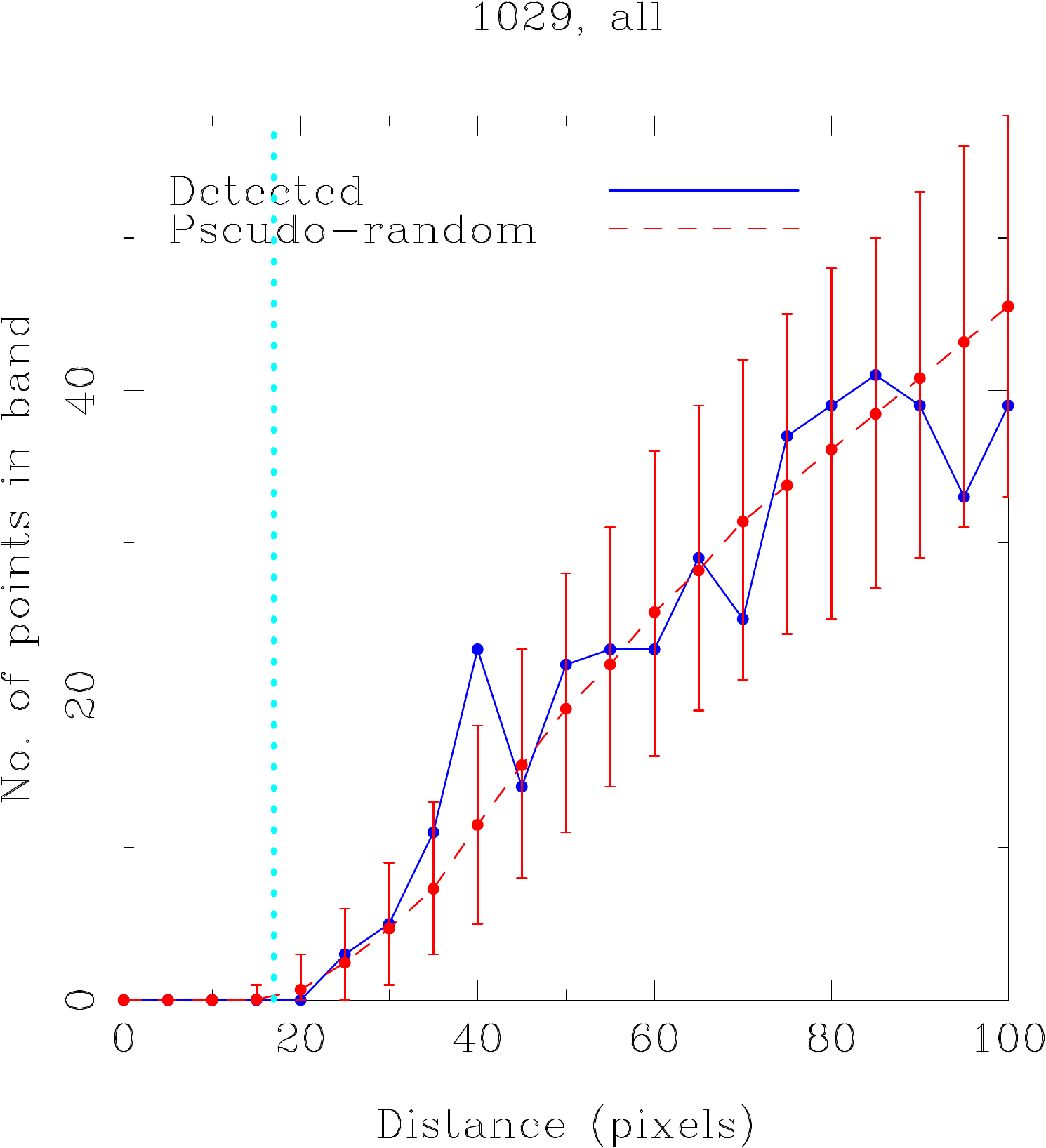}
\hfill \includegraphics[width=0.2\textwidth]{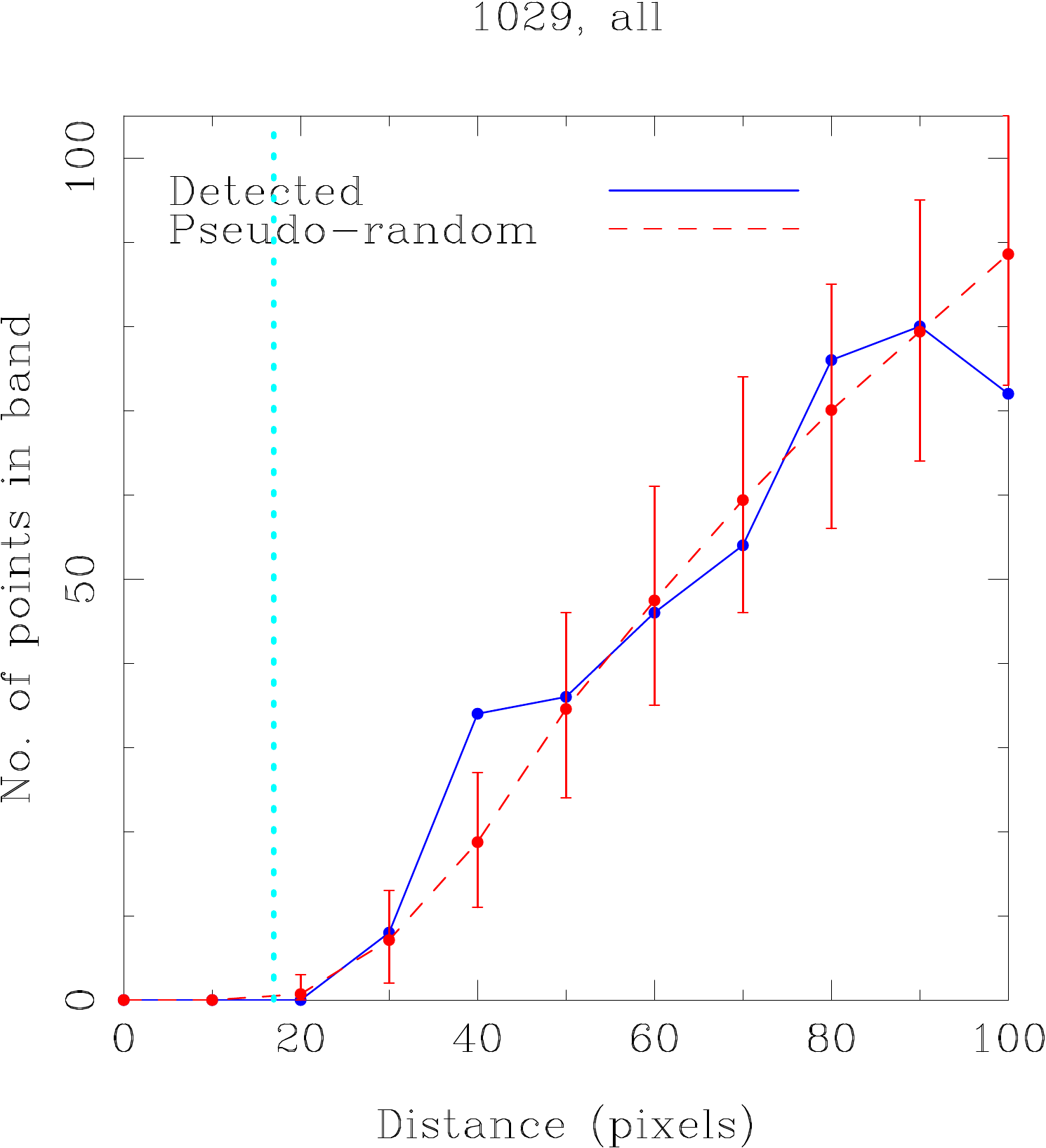}
\hfill \includegraphics[width=0.2\textwidth]{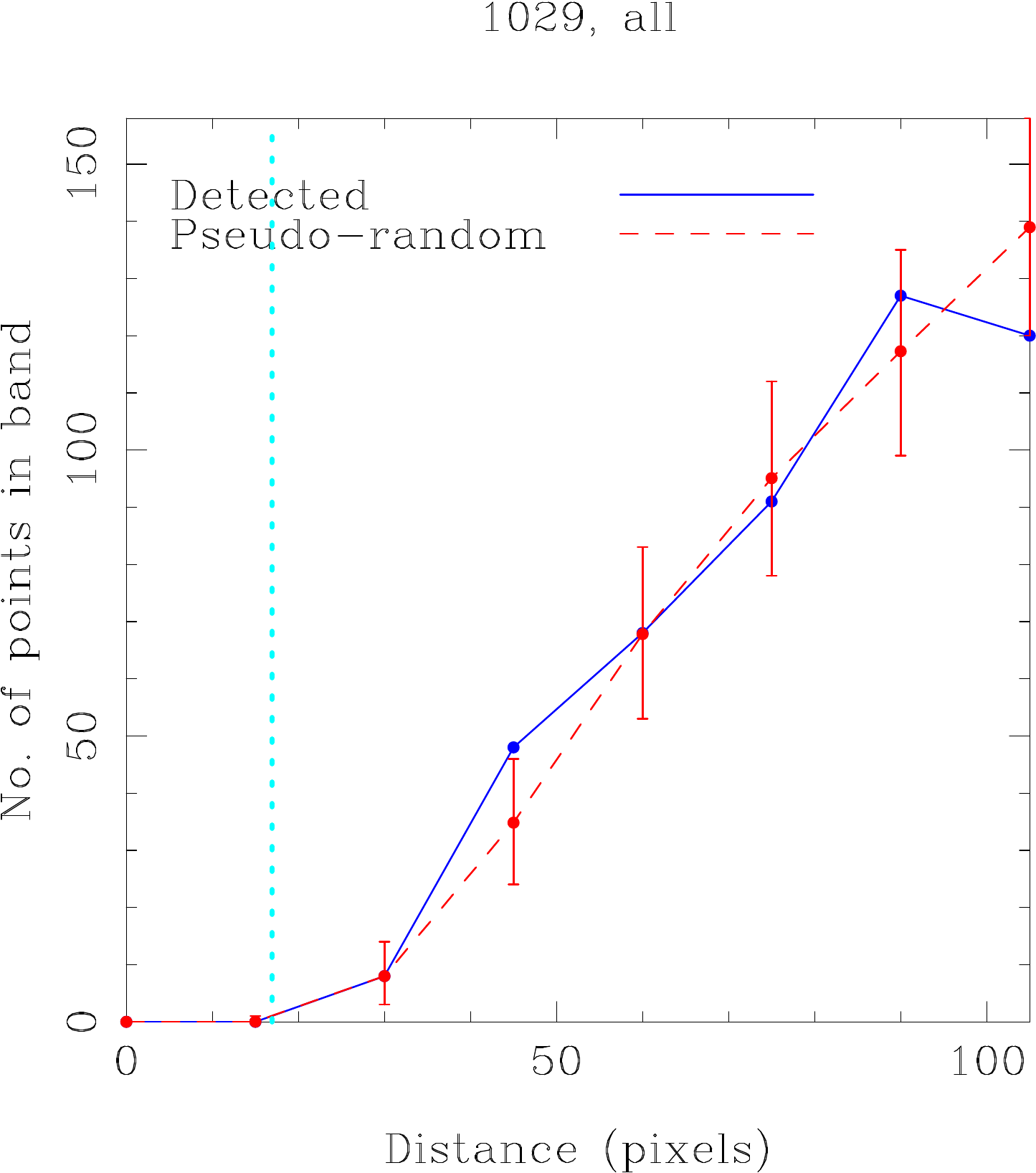}
\caption{Continuation of Figure~\ref{fig:New-features-(short)}.}
\label{fig:New-features-(short)-2}
\end{sidewaysfigure*}

\rotFPbot=0pt

\subsection{Quantitative Tests of Feature Number Enhancement and Suppression}
\label{sec:suppression}
As discussed in Section~\ref{sec:intro}, removal of existing field
and/or suppression of the dynamo mechanism is necessary in order to
prevent the field from growing indefinitely. Suppression of some kind
(including nonlinearity) is necessary to obtain a nonzero steady state
for the field--which is impossible for a strictly linear dynamo
mechanism. Spatially-local suppression would be manifest by an absence in the
occurrence of new features in the proximity to a network concentration
or, equivalently, an enhancement of new features far away. We can
therefore identify the presence or absence of spatially-local suppression by
comparing the appearance of new features with that of the random
result.

For each network concentration, and for each 5-, 10-,
  or 15-pixel-wide annulus around that network concentration, we
produced a histogram of the number of random points placed in each
annulus by the $10^4$ Monte Carlo simulations.  We denote the central
value of the $j$th of $n$ bins of the histogram as $x_j$ and the
histogram value to be $h(x_j)$, such that
$\sum_{j=1}^{n}h(x_{j})=10^{4}$. As a useful check, we observed the
sample $h(x_j)$ to have a distribution that is very close to
Gaussian. We directly measured the mean $\mu$, standard deviation
$\sigma$, and the scaling factor $A$ of the sample, and from those
values derived (not fit) the analytic Gaussian profile of the sample
\begin{equation}
y(x_j)=A\exp(-(x_j-\mu)^2/2\sigma^2)
\end{equation}
We then calculated the reduced $\chi^2$ statistic \citep{Bevington2003},
\begin{equation}
\chi_{\nu}^{2}=\frac{1}{\nu}\sum_{j=1}^{n}\frac{[h(x_{j})-y(x_{j})]^{2}}{h(x_{j})},\label{eq:chi_squared}
\end{equation}
where $\nu=n-n_{c}$ is the number of degrees of freedom, with
$n_{c}=3\,:\,(A,\sigma,\mu)$ being the number of constraints. The
range of the histogram $x_j$'s was truncated so that $h(x_j)>10$ in
order to ameliorate uncertainties in the normality assumption for
small numbers.  $\chi_{\nu}^{2}$ provides a measure of the goodness of
fit of the analytical profile $y(x_{j})$ to the Monte Carlo sample
data $h(x_j)$, and here tends to be between 1 and 2, with most of the
contribution coming from the flanks of the
distribution. $\chi_{\nu}^{2}$ does not contain any information about
the number of detected magnetic features, only how well the
distribution of the number of random points in an annulus approximates a Gaussian.


Using this estimate of the goodness of fit as a guide,
for each network concentration and for each annulus, we determined the
fraction $P$ of the simulations for which the measured number of new
features $N$ in the annulus exceeded the number of random points. This
was done for enhancements above the mean and suppressions below the
mean separately, so that alternating high and low values (indicating
too-small sampling bins) would average out when interpolated to larger
bins. By definition,
\begin{eqnarray}
P= & 1-\frac{\sum_{x_{j}\ge N}y(x_{j})}{\sum_{x_{j}\ge\mu}y(x_{j})}, & N\ge\mu\nonumber \\
P= & -1+\frac{\sum_{x_{j}<N}y(x_{j})}{\sum_{x_{j}<\mu}y(x_{j})}, & N<\mu\label{eq:enhance-suppress-1}
\end{eqnarray}
The interpretation of $P$ is as follows: 
\begin{itemize}
\item A value close to 0 indicates that the number of new features does not 
differ significantly from the number expected from a random distribution; 
\item A value close to $\pm 1$ indicates that the number of new features 
greatly exceeds or is greatly exceeded by that expected from a random 
distribution. 
\end{itemize}
To account for the goodness of fit of $y(x_{j})$ to the sample data,
particularly at small distances where the number of points in an
annulus is small, we scale $P$ at each radius by the corresponding
value of $\chi_{\nu}^{2}$, $Px=\frac{P}{\chi_{\nu}^{2}}$.
Then, we added the values of $Px$ for each of the
  different annuli widths to produce $Px_{tot}$. In this way, we try
  to account for the fact that it is difficult to say what is the
  ``correct'' single annulus width to choose.  The 5-pixel-wide
  annulus is probably too thin and results in large positive
  excursions adjacent to large negative excursions in Figures
  \ref{fig:New-features-(short)} and
  \ref{fig:New-features-(short)-2}. By summing each $Px$ to produce
  $Px_{tot}$, we require that any excursions must be more-or-less
  independent of the resolution of our measurement, which is the
  annulus width.

\begin{figure}
\includegraphics[width=0.8\columnwidth]{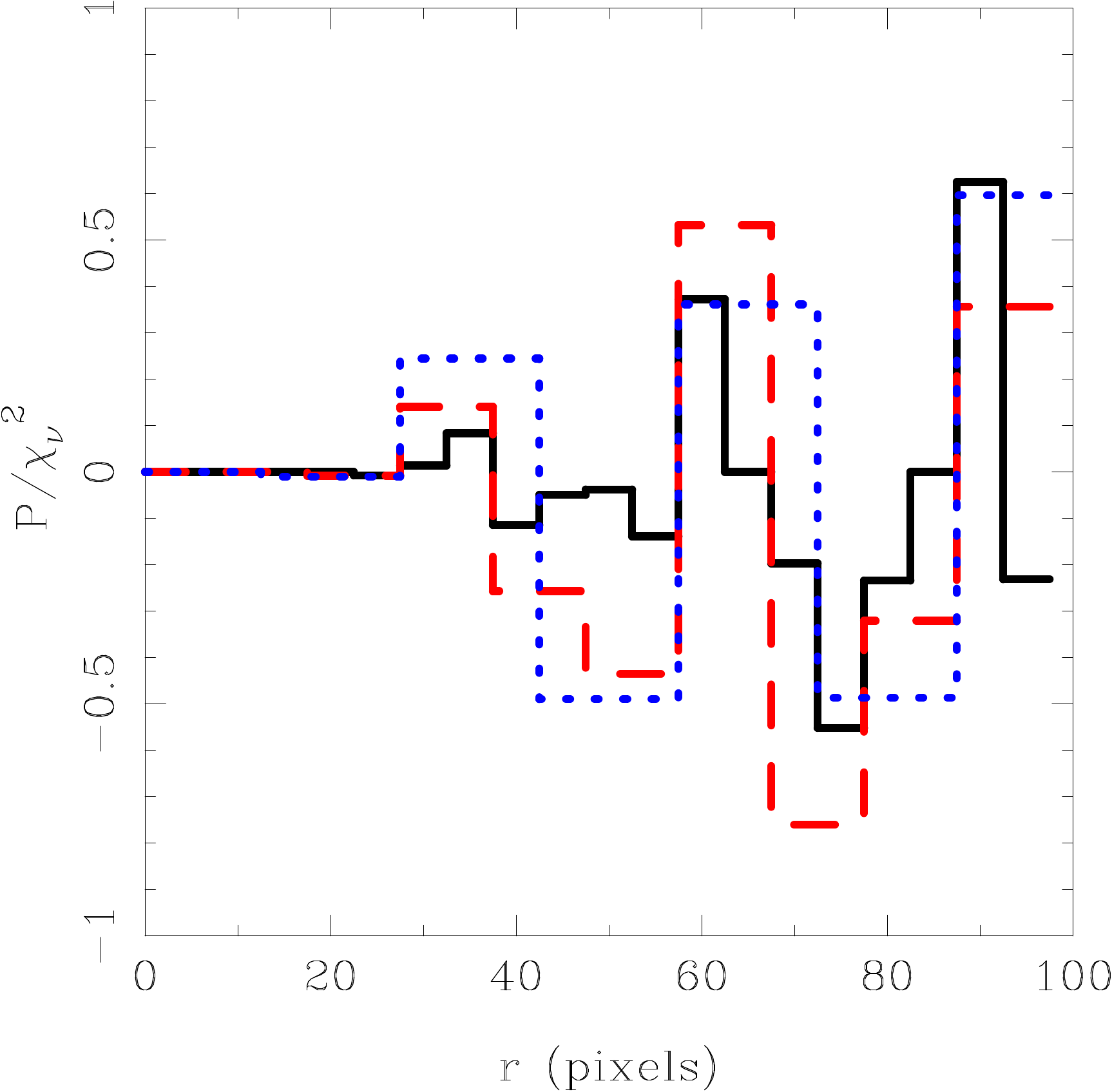}
\caption{For network concentration 153, the quantity
  $\frac{P}{\chi_{\nu}^{2}}$ as a function of distance from the
  network concentration center, for annuli with widths 5 pixels
  (\emph{solid black line}), 10 pixels (\emph{dashed red line}), and
  15 pixels (\emph{dotted blue line}). As described in
    Section \ref{sec:suppression}, a value close to 0 indicates that
    the number of detected features in that annulus does not differ
    significantly from the number exepcted from a random distribution,
    and a large positive (negative) values indicates that the number
    of features is greater (less) than the expected values.}
\label{fig:P-x2-accum}
\end{figure}

For the annulus widths of 10 and 15~pixels we accumulate the values of $P$
into the corresponding 5-pixel bins. Since the values of $Px$ are not
systematically smaller or larger among the three different annulus
widths, a given value in the 15-pixel case transfers directly to the
three corresponding 5-pixel bins, and to the two corresponding bins in
the 10-pixel case. The individual values of $Px$ in each bin were
added algebraically to produce the total $Px_{tot}$ in each 5-pixel
wide bin (Figure~\ref{fig:P-x2-accum}).

As shown in Figure~\ref{fig:P-x2-accum}, $P$ is generally closer to 0
than 1 ($\vert \frac{P}{\chi_{\nu}^{2}}\vert\le 0.6P$) indicating a
statistical preference towards a random distribution, i.e., no
tendency for new features to form away from the network
concentration. We conclude that there is no detectable anomalous spatially-local
suppression taking place.

\subsection{Polarity}
Section~\ref{sec:suppression} demonstrates that the new features 
identified in the present study were not subject to suppression. We 
now explore the mechanism responsible for their enhancement. In 
Section~\ref{sec:Outline}, we described three enhancement 
possibilities, illustrated in Figure~\ref{fig:Shred-&-stretch}: 
shredding, stretching and canceling. Each has a unique measurable 
affect on the polarity of the new features near the network concentration, 
as shown in Table~\ref{tab:NC-mod-polarity-truth}.

\begin{table}
\begin{center}
\caption{Effect of the flux modification of network concentrations 
and their surroundings, and what polarity of new features would be 
expected to be enhanced in the concentrations' surroundings.}
\begin{tabular}{cccc}
\tableline\tableline
 & All & Like & Opposite \\
\tableline\tableline
Shredding & Y & Y & N \\
Stretching & Y & Y & Y \\
Canceling & Y & N & Y \\
\tableline
\label{tab:NC-mod-polarity-truth}
\end{tabular}
\end{center}
\end{table}

We first only considered those features with the same sign as the 
network concentration (a possible signature of either shredding 
or of stretching), and determined their $Px_{tot}^{+}$. Then, we only 
considered those detected features with the opposite sign as the 
network concentration (a signature of canceling or stretching), 
and determined $Px_{tot}^{-}$.

\begin{figure*}
\includegraphics[width=0.3\textwidth]{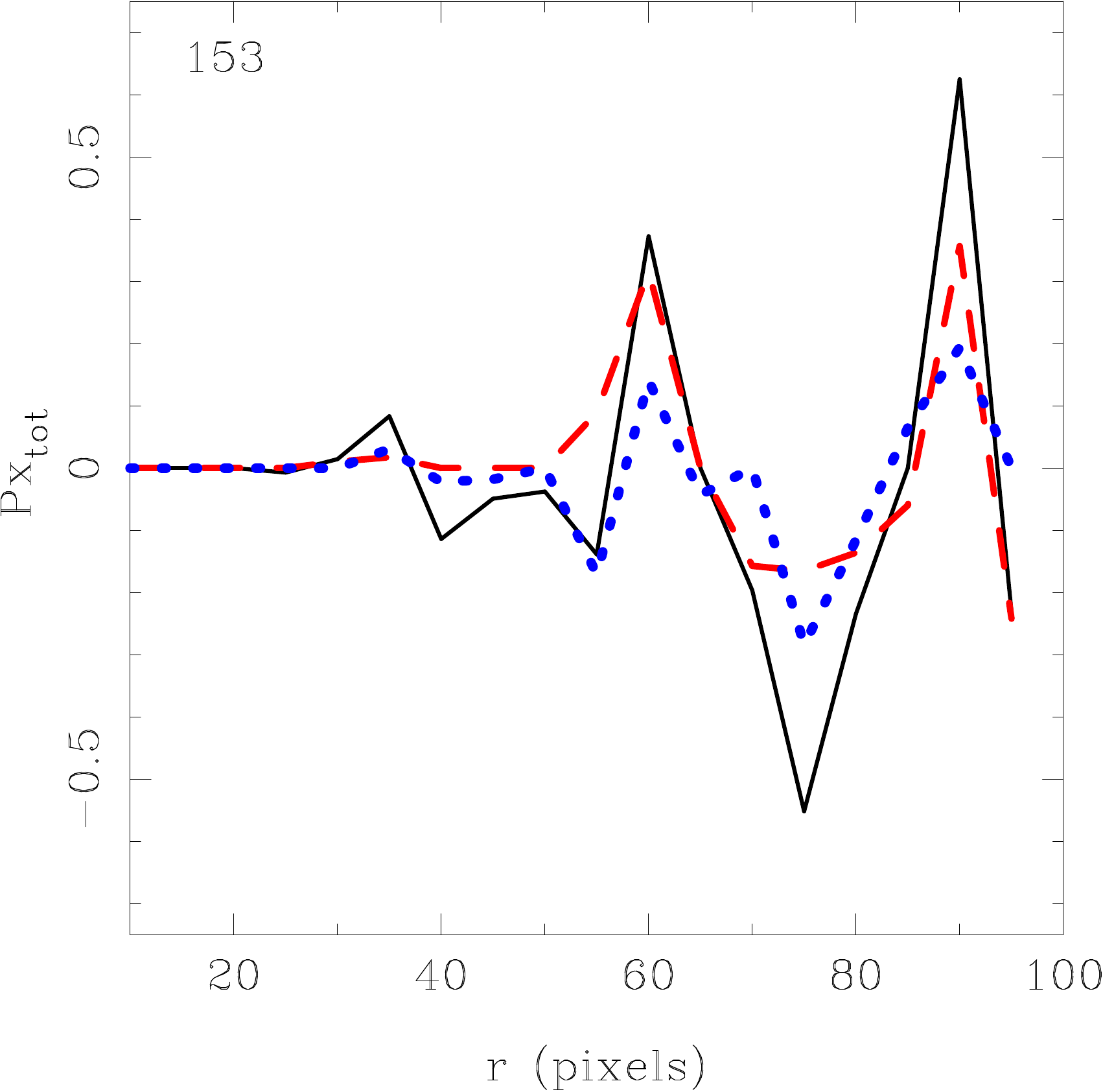}
\includegraphics[width=0.3\textwidth]{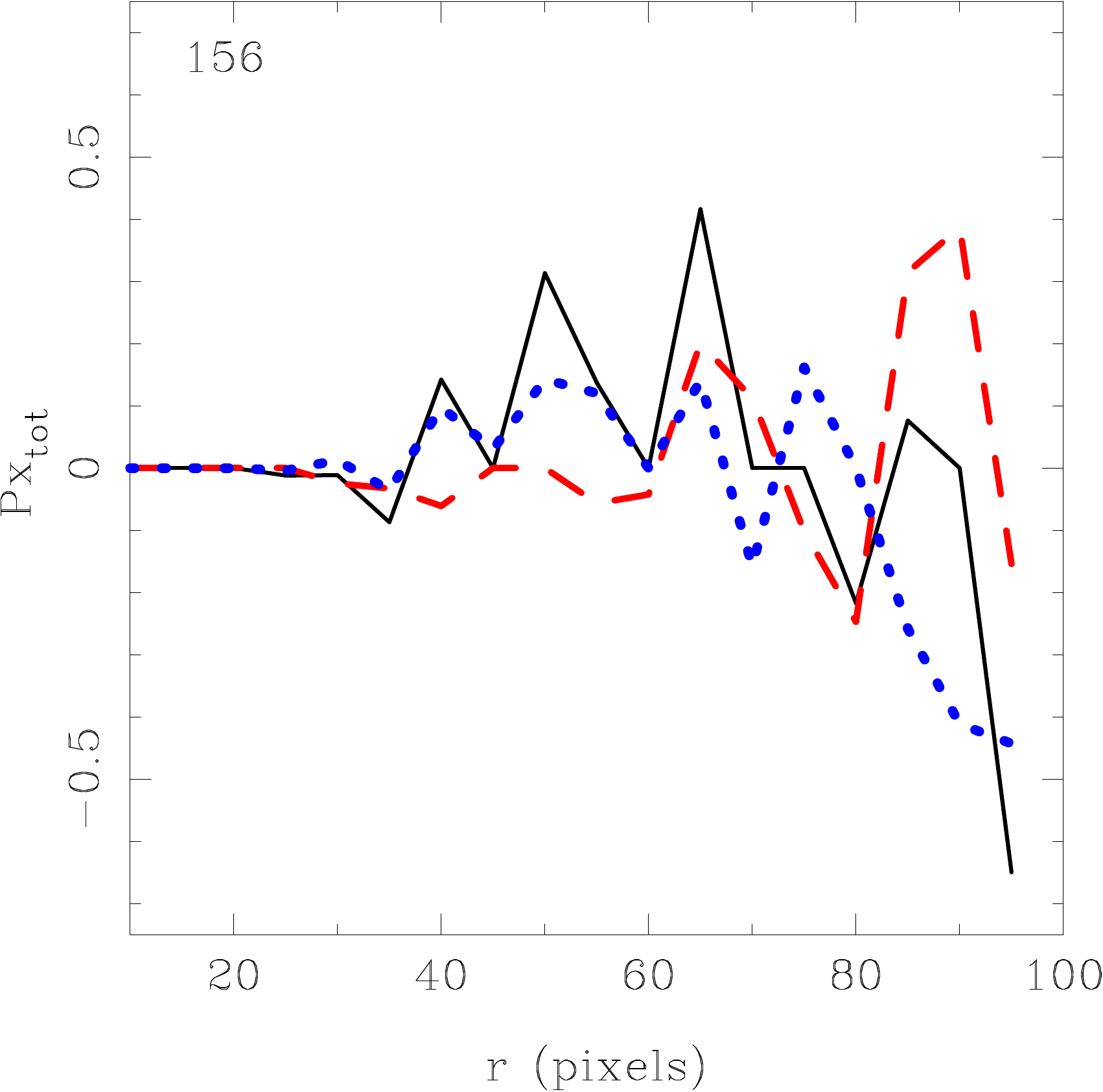}
\includegraphics[width=0.3\textwidth]{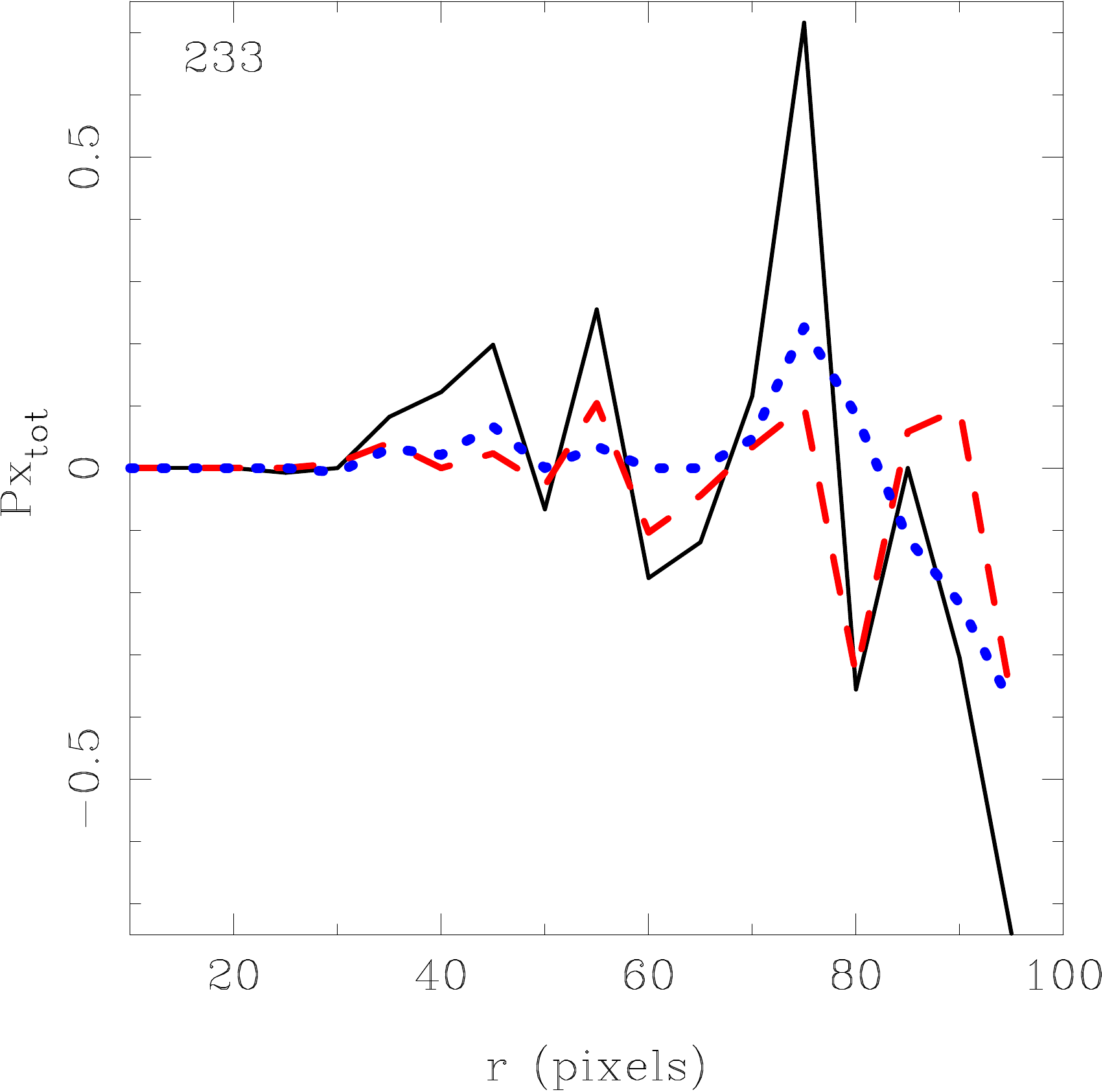}
\includegraphics[width=0.3\textwidth]{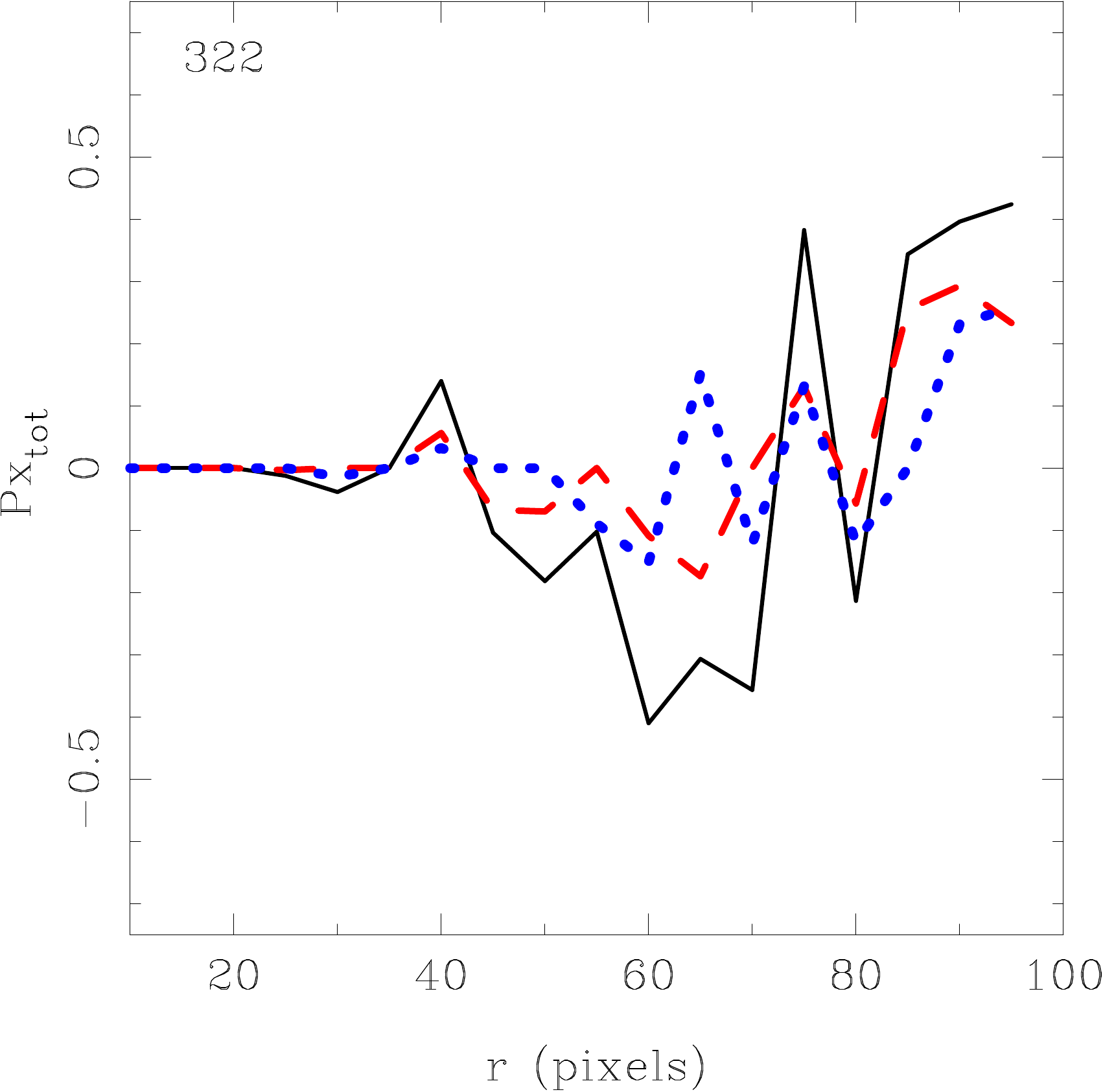}
\includegraphics[width=0.3\textwidth]{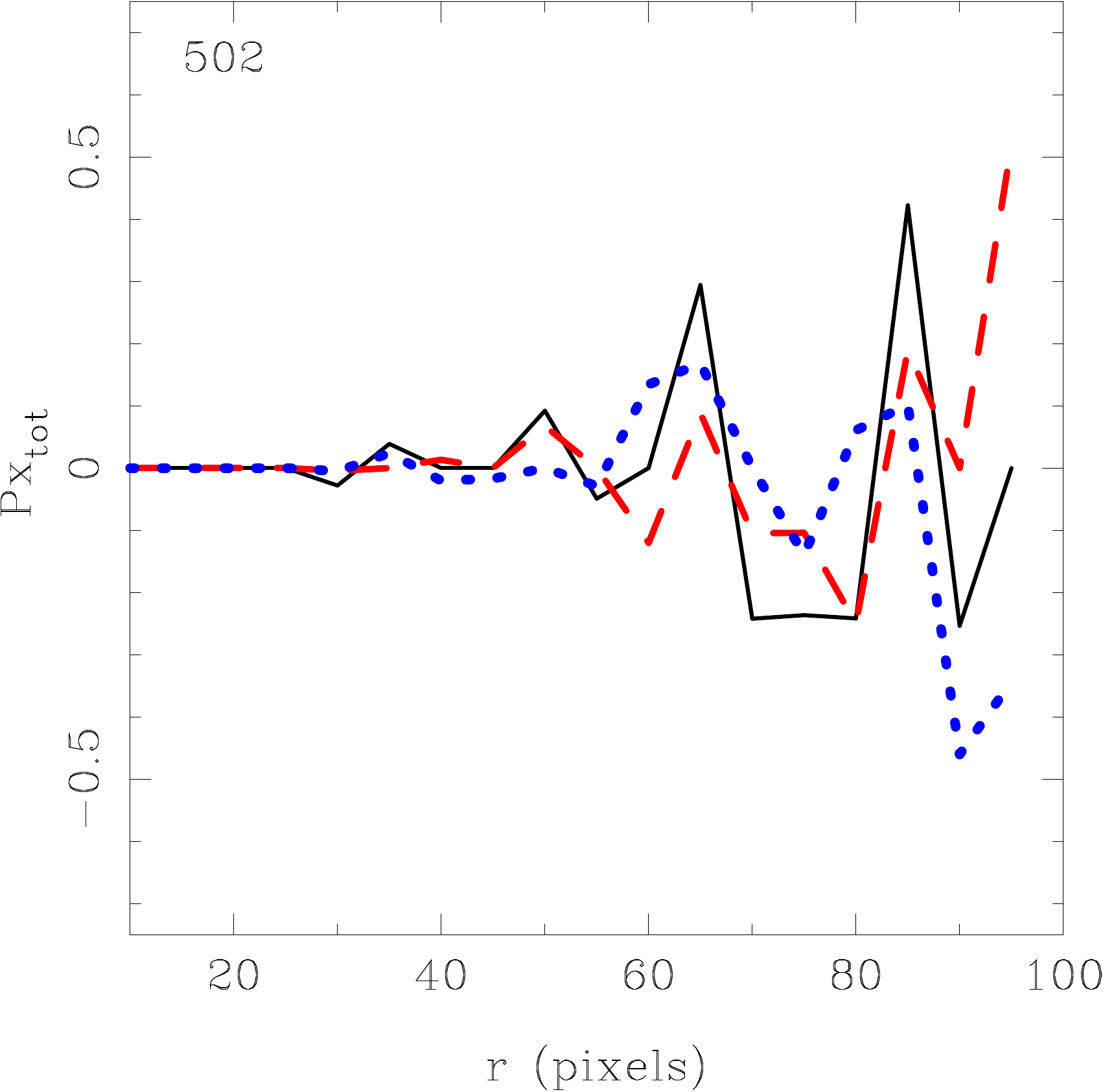}
\includegraphics[width=0.3\textwidth]{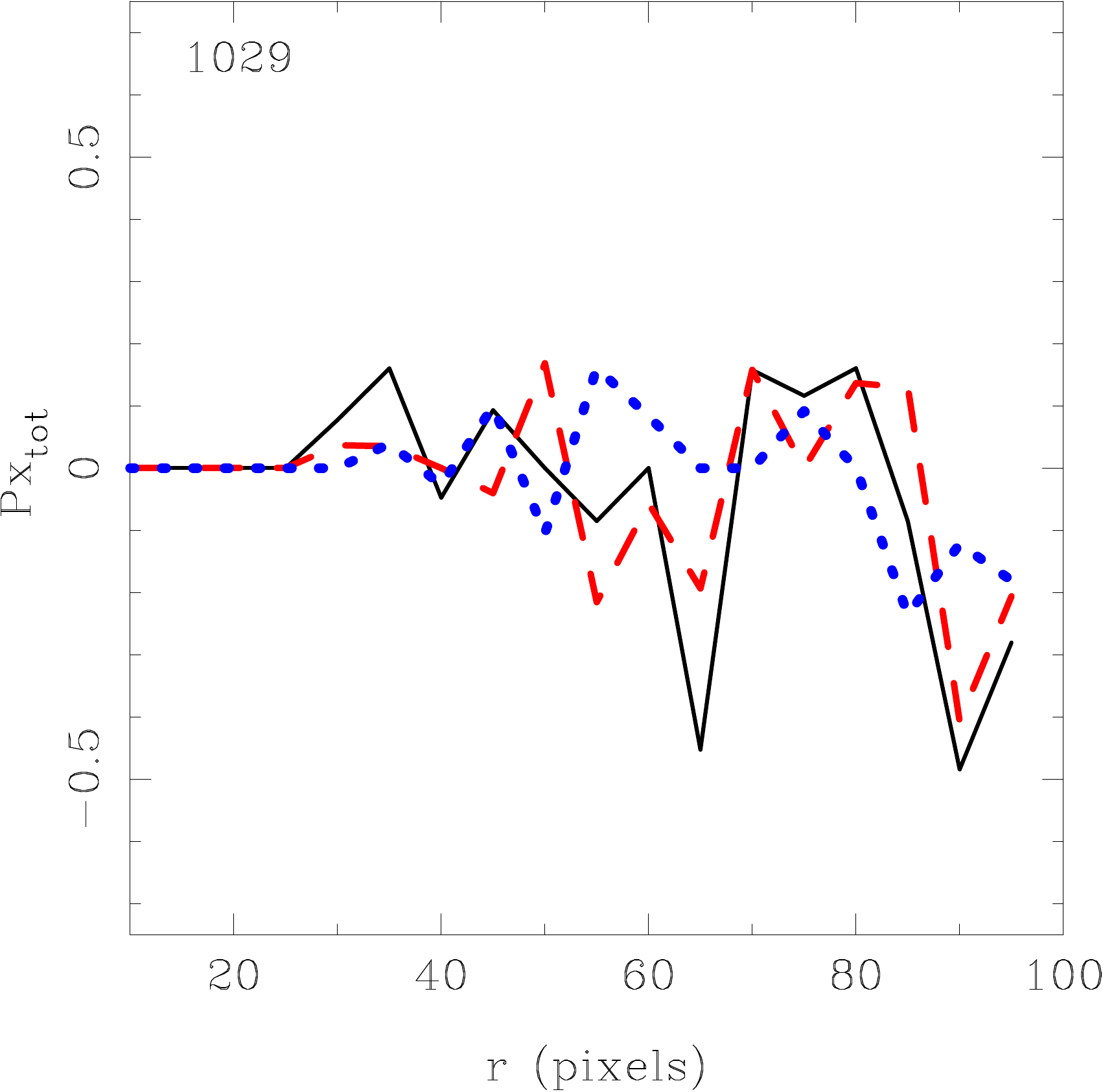}
\includegraphics[width=0.3\textwidth]{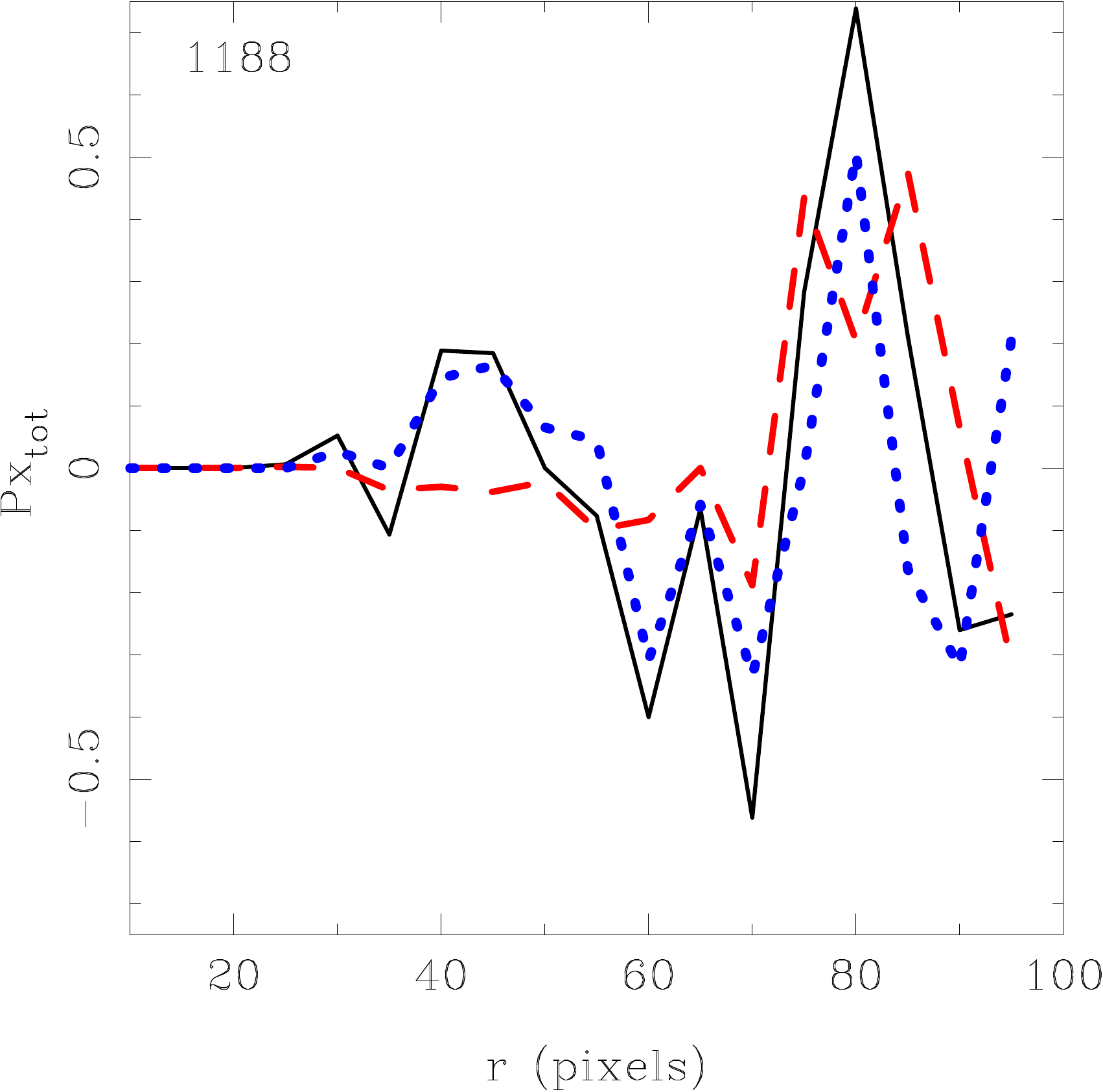}

\caption{Values of $Px_{tot}$ as a function of distance from the network
  concentration center, which indicates the enhancement (for large positive values) or suppression (for large negative values) or lack thereof (for values close to 0) of the number of new features around each network concentration. The three curves correspond to $Px_{tot}$
  (\emph{solid black; all features}), $Px_{tot}^{+}$ (\emph{dashed red; only features of the same polarity as the network concentration}) and
  $Px_{tot}^{-}$ (\emph{dotted blue; only features of the opposite polarity as the network concentration}).}
\label{fig:Px_tot-long}
\end{figure*}

$Px_{tot}(r)$ provides a measure, as a function of distance from the 
center of the network concentration, of the deviation of the number 
of features from the number of random points in an annulus. It takes 
into account the uncertainty in the expected number of points, 
differences in the number of detected features among the network 
concentrations, as well as differences that may arise by changing 
the bin size. Plots of $Px_{tot}(r)$, as well as $Px_{tot}^{+}(r)$ 
and $Px_{tot}^{-}(r)$ for each of the seven network concentrations 
are shown in Figure~\ref{fig:Px_tot-long}.

Since $Px_{tot}$ is a continuous variable, it is useful to have some
indicator of when its value is significantly different from 0. For
annuli widths of 5 pixels, the RMS value of $Px \approx 0.25$.  It is
qualitatively larger for the larger annuli, though the sample size is
small for large annuli.  We sum $Px$ over three annulus sizes (5, 10,
\& 15 pixels), arriving at an overall threshold for $Px_{tot}$ of 0.75
($\approx 2\sigma$).  None of the Figure~\ref{fig:Px_tot-long} network
concentrations have a $Px_{tot}$ that exceeds $\pm0.75$ at any radius.

We do note that for 3 of the 7 network concentrations studied, there
is a peak in the number of features at distances of
$\approx75-90$~pixels.  The peaks are not significant in the
individual concentrations, nor across the ensemble.  Further they are
limited to 1--3~pixel bins in width, and do not have the shape
expected from the mechanisms we discuss. We therefore reject all three
possibilities for a spatially local small-scale dynamo described in
Table~\ref{tab:NC-mod-polarity-truth}.

\section{Network Concentration Evolution}
\label{sub:NetworkConcEvolution}
Our analysis allows us to quantify regions of interest for specific
network concentrations. The results are averages over many things:
azimuthal angle around the network concentration, annuli widths, and
most importantly the 5.25~hour duration of the dataset. In order to
fully understand the interplay between the concentration and the
surroundings, we must also examine the evolution of the network
concentrations themselves.

Many things can affect the evolution of a network concentration. An old 
supergranule may decay, and/or a new one may form, changing the locations 
of the downflow vertex. The downflow may move in response, or it may cease 
to exist entirely, and the flux in the network concentration may be dispersed 
to the lanes of the new supergranule.

The introduction of new flux into an existing flow pattern will also affect 
the network concentration. A strong bipolar region may form in the interior 
of the supergranule, driving flux to the vertex containing the existing 
concentration. Like-polarity flux will increase the strength of the network 
concentration, while opposite-polarity flux will cancel and weaken the 
concentration.

\begin{figure}
\includegraphics[width=1.0\columnwidth]{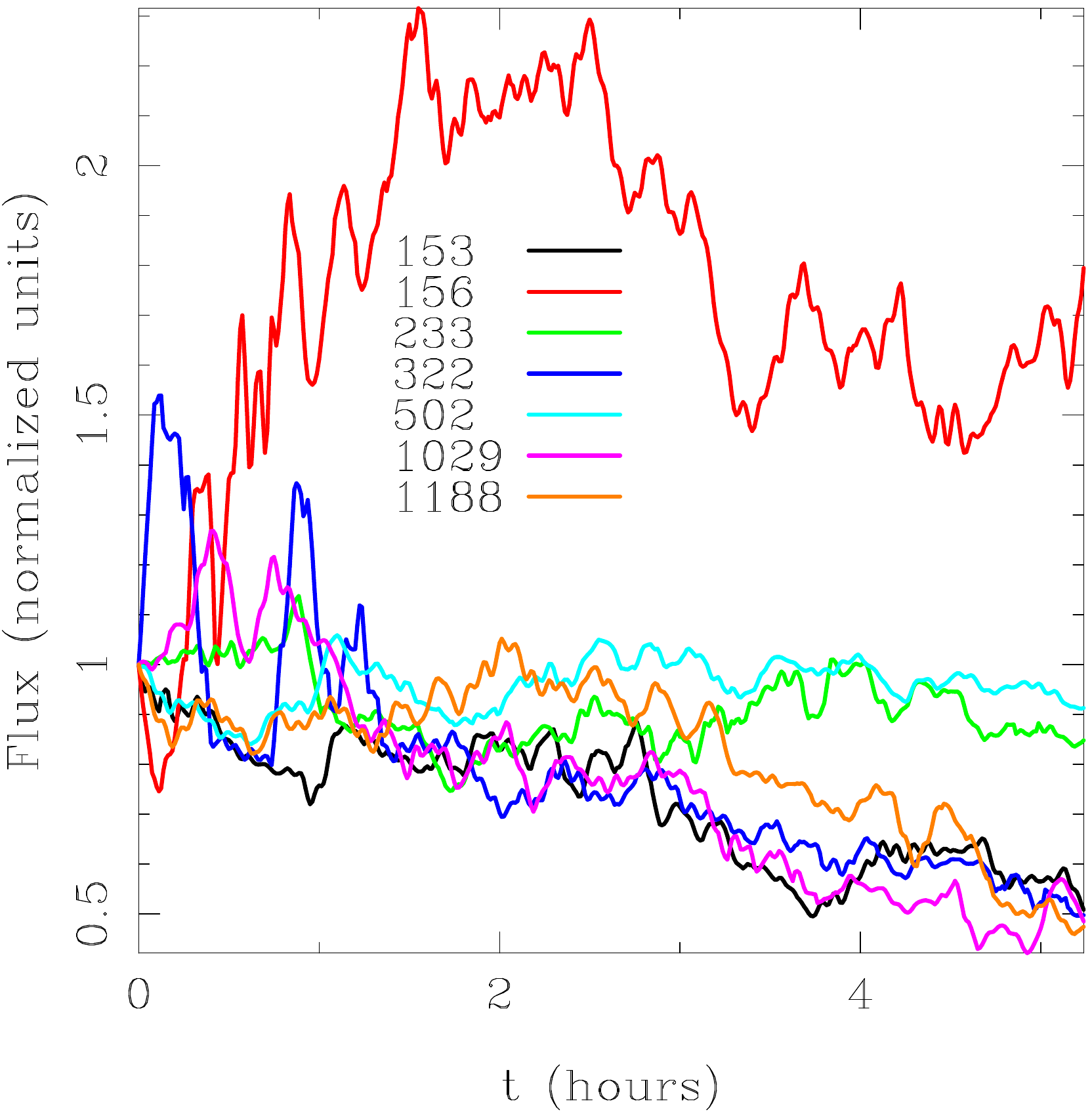}
\caption{Evolution of network concentration flux.
The lines shown here have been convolved with a 10-frame boxcar function
to smooth out short term variations, and then normalized by their
initial values.}
\label{fig:Evol-NC-flux}
\end{figure}

Figure~\ref{fig:Evol-NC-flux} shows the flux magnitude evolution of each of 
the seven network concentrations across the dataset. Each curve has been convolved 
with a 10-frame boxcar kernel to smooth out short-term variations, and then normalized 
by its initial value. Several of the network concentrations lose large 
fractions of their flux over the 5.25 hours; three lose more than 50\%. 
Table~\ref{tab:NC-Properties} describes various evolution properties of the 
network concentrations and their surroundings. Note that edge effects due 
to the convolution cause some concentrations' final fluxes to appear slightly 
different in Figure~\ref{fig:Evol-NC-flux} than reported in 
Table~\ref{tab:NC-Properties}. 

\begin{table*}
\begin{center}
\caption{Properties of selected network concentrations
and their surroundings. The network concentrations' flux ($\Phi$)
and flux density ($\langle B\rangle$), and the surroundings' flux
imbalance ($\xi$) and unsigned flux density ($\langle|B|\rangle$),
are all medians taken over the length of the dataset. The network
concentrations' change in flux $\frac{\Phi_{F}}{\Phi_{I}}$ uses the
raw, unsmoothed data.}
\begin{tabular}{ccccccc}
\tableline\tableline
\multicolumn{4}{c}{Network Concentration} & \multicolumn{2}{c}{Surroundings} & \\
ID & $\Phi(10^{18}\textrm{ Mx})$ & $\langle B\rangle$ (G) & $\frac{\Phi_{F}}{\Phi_{I}}$(\%) & $\xi$ (\%) &  $\langle\left|B\right|\rangle$ (G) & Evolution Summary \\
\tableline\tableline 
N/A & N/A & N/A & N/A & 12.1 & 7.91 &  (Dataset taken as a whole)\\
153 & 16.7 & -97.8 & 48 & 26.0 & 12.0 & Nearby opp. polarity, but NC shredding \\
156 & 9.75 & -93.4 & 174 & 28.3 & 11.5 & much merg/frag to make a distinct NC \\
233 & 32.9 & -127 & 84 & 17.5 & 9.58 & much merg/frag to make a distinct NC \\
322 & 18.5 & -110 & 59 & 47.9 & 8.49 & steadily cleaving off field \\
502 & 90.2 & -158 & 92 & -39.0 & 8.85 & very large NC, not much change \\
1029 & 10.1 & +104 & 47 & 47.9 & 15.4 & shredding trailing side, cancel leading side \\
1188 & 23.7 & +103 & 45 & -42.5 & 9.23 & NC cancellation \\
\tableline
\label{tab:NC-Properties}
\end{tabular}
\end{center}
\end{table*}

Movies of the network concentration evolution suggest that the
dynamics of the concentrations themselves are likely to dominate the
surroundings, and other quantities listed in
Table~\ref{tab:NC-Properties} confirm this idea.  For example,
Concentration 1188 shows a 55\%\ decrease in flux over 5.25 hours, and
has an enhancement of detected features due to features of polarity
opposite to itself. The flux imbalance of the surroundings is defined as
the ratio of the net flux to the absolute
  flux \citep[see][Eq. 1]{Hagenaar2008}
\begin{equation}
\xi \equiv S\times\frac{\Sigma\Phi_{i,j}}{\Sigma|\Phi_{i,j}|},\label{eq:flux-imbalance}
\end{equation}
where $S$ is $\pm1$ for positive/negative network concentrations,
respectively, and $\Phi_{i,j}$ is the flux in pixels outside the
concentration's maximum spatial extent. In the region surrounding
network concentration (NC) 1188, $\xi$ is --43\%, confirming the
excess of opposite polarity flux in the region. A movie shows that
this concentration experiences a large inflow of opposite polarity flux,
which cancels the flux in the parent concentration. NC 1029, on the
other hand, is more complex. A movie shows that the concentration is
in motion, canceling with a cluster of opposite polarity flux on the
leading edge, while shredding like-polarity flux off the trailing
edge. These processes cause a decrease in the concentration's flux of
52\%. Around NC 153 there is a strong opposite polarity cluster of
flux nearby, but the evolution of the NC itself appears to be
dominated by shredding of 53\%\ of its flux.

The other network concentration that exhibits a large decrease in flux (nearly 
40\%), is NC 322, which shows a reduction of new features at a distance of 
60--70~pixels. There is no polarity preference to this reduction, so one 
cannot infer that it is due to the cleaving of flux off the network 
concentration, nor to small features canceling the network concentration's 
flux. Rather, there seems to be a ``halo'' around the network concentration 
within which few weak, small features of either polarity are born. We believe 
that this may be due to suppression, and is the only possible sign of 
suppression of new features among the seven network concentrations studied.

In contrast, the flux of NC 233 changes by only 15\%\ over 5.25~hours.
There is no large suppression of features except at large distances
(90--100~pixels). This concentration exhibits the classic signatures
of being at the intersection of three supergranular lanes. Towards the
beginning of the dataset, a portion of the network concentration flux
is shredded off towards the north, but simultaneously another cluster
of like-polarity flux joins the concentration from the
southeast. Throughout the time series, opposite polarity flux flows
into the concentration from all directions and cancels.

In short, the small excursions in feature count observed near these
flux concentrations can be understood in terms of the evolution of the
features as they interact with measurable patterns in their
surroundings. This strengthens our null result, that there is no
significant \emph{spatially local small-scale} dynamo action, because these small excursions
observed in the Figure~\ref{fig:Px_tot-long} distributions are themselves consistent with
confounding processes. These processes are inside the statistical
noise floor of the present analysis, but must be taken into account in
any attempt to further increase the sensitivity of the null
measurement.

\section{Discussion}
\label{sec:Clustering-Discussion}
The objective of our work was to test various solar dynamo models
using observations. This was made possible by automated feature
tracking techniques, which enabled the identification of small, weak
magnetic features in the vicinity of large, evolving network
concentrations. We treated the number of these small weak features as
a proxy for the vigor of the flux production process.

Through the statistical analyses detailed in Section~\ref{sec:results} we
have found, with our sample of seven network concentrations observed
with \emph{Hinode}/NFI, that small-scale magnetic features form in a
random distribution at least within our observable range of
$\sim$12~Mm from the center of the network concentrations. We found no
observable tendency for newly-formed features to either cluster around
the neighborhood of network concentrations, nor to have their
formation rate suppressed there. We conclude that there is no
spatially local small-scale dynamo action due to stretching and
subsequent emergence of nearby subsurface fields, nor suppression of
small-scale dynamo action due to the presence of nearby strong field.

Comparing the polarity of newly-formed features to that of the nearby
network concentration allowed us to probe the mechanisms of shredding
\citep{Schrijver1997} independent from a hypothetical small-scale dynamo.
The lack of a spatial locality signature in the polarity of the new feature
distribution indicates that shredding plays at most a very minor role
in the flux balance of the nearby network.  We draw the conclusion
that there is no influence of small-scale magnetic feature
enhancement from a nearby network concentration. 

The last two results together rule out every major possibility for a
dynamo that is local on these spatial scales, i.e., one that works by
stretching or recycling nearby flux, and is therefore dependent on the
strength of fields in the neighborhood rather than the global volume
of the convection zone.

\begin{figure*}
\includegraphics[width=1.0\textwidth]{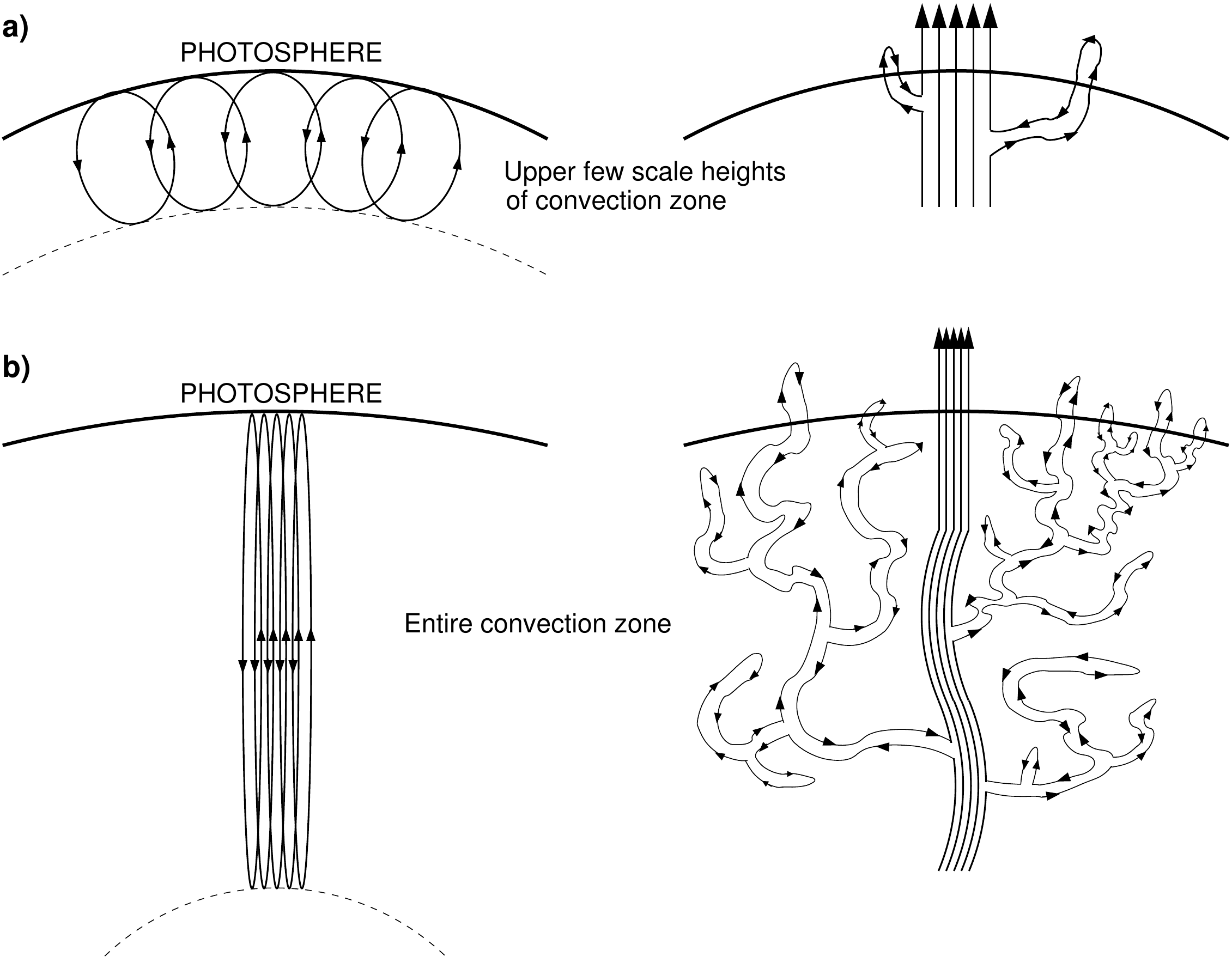}
\caption{Illustrations of the two basic categories of model describing
  the nature of the small-scale solar dynamo. The left column provides an
  elementary illustration of the category, and the right column
  illustrates a scenario in which the sub-surface field beneath a
  network concentration can become distorted and breach the
  photosphere elsewhere. a) The dynamo extends to small depths, down
  to the upper few scale heights of the convection zone. In this case,
  parts of the sub-surface field beneath a network concentration is
  most likely to breach the photosphere at a location near the
  concentration itself. b) The dynamo extends to large depths, through
  the entire convection zone.  In this case, the breaching of the
  photosphere by the sub-surface field does not necessarily occur near
  the seed network concentration.}
\label{fig:two-dynamo}
\end{figure*}

We now return to the two types of solar dynamo mentioned in
Section~\ref{sec:intro}: shallow and deep. These are illustrated in
the left column of Figure~\ref{fig:two-dynamo}. Our analysis of the
location of feature birth rules out local suppression, and our
analysis of the polarity of the features excludes shredding and
canceling of the network concentration. Stretching at the local level
is also excluded, as this would require newly-formed magnetic features
to be in relative close proximity to the network concentration, as it
would be unlikely that a small sub-surface field could be distorted to
any large distance from the concentration itself (see
Figure~\ref{fig:two-dynamo}a). Our main conclusion therefore is that
there is no spatially local dynamo
  on the spatial scales observable by Hinode, i.e. that the observed
  small-scale field is a high- wavenumber manifestation of a large-
  scale phenomenon. We note that all plausible surface dynamo models
  have the property of spatial locality, and we are therefore able to
  exclude the possibility of a surface dynamo.

This leaves us with the second category, a deep small-scale dynamo,
illustrated in Figure~\ref{fig:two-dynamo}b as suggested by
\cite{Stein2003,SteinNordlund2006}. A deep dynamo would allow for
stretching at much larger distances, and so there would be no preference
for small-scale enhancements in the proximity of the network
concentration. There would also be no preference for polarity in the
proximity of the network concentration due to the random tendency of
the subsurface flows to breach the photosphere.

\subsection{Limitations of the Result}
In many cases, the network concentration appeared to be affected 
by nearby small-scale magnetic features. This implies that weak fields
coming off or onto the network concentration can be just at the limits
of visibility in the NFI magnetograms and still have an effect on
the parent network concentration. While the effect of a single one
of these features is small, this affect from a large number may play
an important role in the network concentration evolution.

It is important to call attention to the limits of our selected
parameter $P$ for comparing the new features with those of the random
points.  Firstly, even with a purely random distribution, we should
not expect the integral of $Px_{tot}$ in Figure~\ref{fig:Px_tot-long}
to be zero, because of the uncertainties associated with the factors
that went into its derivation. Secondly, $P$ is not effective at
identifying polarity concentrations on opposite sides of a network
concentration, since it averages over many variables. This is best
demonstrated with a movie of NC 1029. If a network concentration had a
surplus of like polarity features on one side (perhaps due to
shredding) and a surplus of opposite polarity features on the other
side (perhaps due to cancellation), and these features happened to be
at or near the same distance from the center of the network
concentration, then $Px_{tot}$, $Px_{tot}^{+}$ and $Px_{tot}^{-}$
could all be enhanced at the same distance, which would mimic the
signature of subsurface field line stretching and emergence. It is
therefore necessary to accompany the analysis $Px_{tot}$ with at least
a visual inspection of the evolution of the network concentration (the
classic balance of case studies vs.\ statistics).

It is important to reiterate that the features that one readily sees
merging into and fragmenting from the network concentrations are not
the same population as those considered in our analysis. This is
because we have excluded those features born by
``Fragmentation'' and ``Error'' (see Section
  \ref{sub:Position-New-Features}), which accounts for a large
fraction of the features in the dataset, and an even larger fraction
of the strong ones \citep[see][]{Lamb2010,Lamb2013}.  Analyzing only
these small, weak features allows us to consider only the true
small-scale field as observed by \emph{Hinode}/NFI.

\subsection{Concluding Remarks}
We have conducted a statistical analysis of the birth and polarity of
small-scale magnetic features in proximity to strong supergranular
network concentrations. We find no observational evidence of a
relationship between the network concentration and the
number or polarity of features at a given distance
  from the network concentration. This rejects the
spatially local small-scale solar dynamo hypothesis at
  spatial scales observable by \emph{Hinode}/NFI, and we therefore
  conclude that the features that dominate the photospheric magnetic
  landscape are driven by a spatially nonlocal deep solar dynamo.

\acknowledgements 

We thank Richard Frazin and the anonymous referee for comments which
improved the paper. The authors were partially supported by NASA
Grants NNX08AJ06G and NNX11AP03G. We acknowledge the Magnetic Feature
Tracking Workshops, support for which was provided by NASA's SHP/GI
program. We thank the \emph{Hinode} team for making their data
publicly available. \emph{Hinode} is a Japanese mission developed and
launched by ISAS/JAXA, with NAOJ as domestic partner and NASA and STFC
(UK) as international partners. It is operated by these agencies in
co-operation with ESA and NSC (Norway).

\end{document}